\documentclass{nature}

\usepackage{amsmath}
\usepackage{graphicx}
\usepackage{xfrac}
\usepackage{multirow}
\usepackage{subfigure}
\usepackage{mathtools}
\usepackage[colorlinks=true,linkcolor=blue,citecolor=black,urlcolor=blue]{hyperref}
\usepackage{soul}
\usepackage[usenames,dvipsnames]{xcolor}
\usepackage{authblk}
\usepackage[labelfont=bf]{caption}
\usepackage[figurename=Figure ]{caption}
\usepackage{siunitx}
\usepackage{scalefnt}

\usepackage{lineno}

\makeatletter

\let\saved@includegraphics\includegraphics
\AtBeginDocument{\let\includegraphics\saved@includegraphics}

\makeatother

\usepackage{color}
\usepackage{dsfont}
\usepackage[normalem]{ulem}
\usepackage{dcolumn}
\usepackage{pdfpages}
\usepackage{blindtext}
\usepackage{outlines}
\usepackage{enumitem}
\usepackage{soul}
\setlist[enumerate,2]{label=\roman*)}
\setlist[enumerate,3]{label=\alph*)}
\usepackage{multirow}

\renewcommand{\vec}[1]{\mathbf{#1}}

\DeclareSIUnit\mub{\mu_\text{B}}

\begin{document}

\title{Non-Majorana modes in diluted spin chains proximitized to a superconductor}

\author{Felix K{\"u}ster$^{1}$, Sascha Brinker$^{2}$, Richard Hess$^{3}$, Daniel Loss$^{3}$, Stuart S. P. Parkin$^{1}$, Jelena Klinovaja$^{3}\ast$, Samir Lounis$^{2,4\ast}$, Paolo Sessi$^{1\ast}$}
\affil{$^1$Max Planck Institute of Microstructure Physics, Halle 06120, Germany\\
$^2$Peter Gr{\"u}nberg Institut and Institute for Advanced Simulation, Forschungszentrum J{\"u}lich \& JARA, J{\"u}lich D-52425, Germany\\
$^3$Department of Physics, University of Basel, Klingelbergstrasse 82, CH-4056 Basel, Switzerland\\
$^4$Faculty of Physics, University of Duisburg-Essen and CENIDE, 47053 Duisburg, Germany}
\affil{$^\ast$ Emails: jelena.klinovaja@unibas.ch, s.lounis@fz-juelich.de, paolo.sessi@mpi-halle.mpg.de}

%\date{\today}	

\maketitle

\vspace{1cm}

\begin{abstract}
Spin chains proximitized with superconducting condensates have emerged as one of the most promising platforms for the realization of Majorana modes. Here, we craft diluted spin chains atom-by-atom following seminal theoretical proposal suggesting indirect coupling mechanisms as a viable route to trigger topological superconductivity.  Starting from single adatoms hosting deep Shiba states, we use the highly anisotropic Fermi surface of the substrate to create spin chains characterized by different magnetic configurations along distinct crystallographic directions. By scrutinizing a large set of parameters we reveal the ubiquitous emergence of boundary modes. Although mimicking signatures of Majorana modes, the end modes are identified as topologically trivial Shiba states. Our work demonstrates that zero-energy modes in spin chains proximitized to superconductors are not necessarily a link to Majorana modes while simultaneously identifying new experimental platforms, driving mechanisms, and test protocols for the determination of topologically non-trivial superconducting phases.
\end{abstract}

%\vspace{1cm}

\section*{Introduction}

The development of topological concepts in condensed matter systems has motivated much interest in the realization of Majorana states \cite{Alicea_2012,Beenakker}. These exotic states are predicted to emerge at boundaries of topological superconductors \cite{Sato_2017}, manifesting themselves as zero-energy modes in conductance measurements \cite{Mourik1003,Nadj-Perge602,PhysRevLett.115.197204,PKK2016,Kimeaar5251,KCW2018,PhysRevLett.114.017001}. Beyond their fundamental interest, which illustrates how topological condensed matter systems can be extremely fertile in establishing strong connections with concepts developed in the world of high energy physics \cite{Majorana,Wilczek2009}, the associated non-Abelian exchange statistics has raised great expectations for their direct application in topological quantum computational schemes \cite{AOR2011}. Following Kitaev's seminal proposal \cite{Kitaev_2001}, several distinct platforms have been theoretically proposed for their experimental realization, most notably the proximitization of a conventional $s$-wave superconductor to allow for spin-split states, such as by hosting them on topological insulators \cite{PhysRevLett.100.096407}, semiconductors with strong spin-orbit coupling \cite{PhysRevLett.105.077001,PhysRevLett.105.177002}, or via magnetic nanostructures \cite{PhysRevB.88.020407,PhysRevLett.111.186805,PhysRevB.88.155420}. 

Since Majorana fermions are a direct manifestation of a topologically non-trivial superconducting state, they are expected to reveal themselves as boundary excitations. For these reasons, spatially resolved spectroscopic techniques are especially suitable to atomically resolve the emergence of Majorana states and to disentangle them from spurious and topologically trivial zero-bias states \cite{Nadj-Perge602,PhysRevLett.115.197204,PKK2016,PhysRevLett.114.017001,KHV2020}. In this context, scanning tunneling microscopy and spectroscopy measurements play a critical role. Following earlier works on self-assembled magnetic chains \cite{Nadj-Perge602,PhysRevLett.115.197204,PKK2016}, the recent use of atomic manipulation techniques has allowed for the construction of disorder-free chains, which has provided insights into the creation and the manipulation of Majorana states \cite{Kimeaar5251,KCW2018,SBS2020,SBP2021,PhysRevB.104.045406}. 

For the building blocks, several theoretical proposals have highlighted the important role of having diluted impurities with the superconducting host playing an active role in mediating their coupling~\cite{PhysRevB.88.020407,PhysRevLett.111.186805,PhysRevB.88.155420}. Despite the appealing prospects of the proposed topological concepts and scenarios, their experimental realization has remained elusive and largely unexplored. Indeed, all systems explored so far are densely packed and ferromagnetically coupled through direct magnetic exchange~\cite{Nadj-Perge602,PhysRevLett.115.197204,PKK2016,Kimeaar5251,KCW2018,SBS2020,SBP2021,PhysRevB.104.045406}.

Here, we use atomic manipulation techniques to create dilute spin chains, where Shiba bands are generated by indirect coupling mediated by the superconducting condensate~\cite{Dinge2024837118,kuster2021long}. 
We demonstrate that the presence of Yu-Shiba-Rusinov (YSR, or Shiba) pairs very close to the center of the superconducting gap allows the emergence of modes highly localized at the ends of the chains for a large variety of parameters such as coupling strengths, magnetic configurations, and hopping amplitude, and distinct unit cell. Despite displaying some spectroscopic signatures compatible with Majorana bound states, the end modes are identified as topologically trivial Shiba states energetically located at or very close to zero.

\section*{Results}

\subsection{Creation of zero-energy boundary modes}

One-dimensional spin chains are created atom-by-atom utilizing atomic manipulation in a STM operated at cryogenic temperatures. Details on sample preparation and measurement protocols can be found in Methods. We focus on Cr atoms indirectly coupled through the (110) surface of niobium, a system which has recently been demonstrated to allow for the tuning of interactions by actively using the  superconducting condensate to mediate indirect coupling among the localized Cr spins \cite{kuster2021long}. 
By contrast with atomic chains created by self-assembly, the use of direct atomic manipulation allows for much greater flexibility since it makes it possible to vary several key experimental parameters such as: (i) chain length, (ii) crystallographic direction, and (iii) distance between local magnetic moments. Despite considerable progress in the field, earlier studies focused on densely packed spin chains, where atoms are nearest neighbors and their coupling is dominated by ferromagnetic direct magnetic exchange \cite{Nadj-Perge602,PhysRevLett.115.197204,PKK2016,Kimeaar5251,KCW2018,SBS2020,SBP2021,PhysRevB.104.045406}. 

Figure 1 reports the creation and spectroscopic characterization of spin chains built atom-by-atom along the crystallographic direction $[1\overline{1}3]$. As illustrated in panel a, the distance between successive ad-atoms amounts to \SI{0.55}{\nano\meter}. Figure 1b illustrates the spectroscopic characterization of the atomically crafted chains for representative chain lengths up to 10 atoms. All spectroscopic measurements have been acquired using a superconducting tip. This allows to investigate the electron–hole symmetry of the chain's zero-energy modes, which correspond to the tip superconducting energy gap $\pm\Delta_\mathrm{tip}$ (see black dashed line). More spectroscopic data can be found in Supplementary Notes 1-3. 

\begin{figure*}
        \renewcommand{\figurename}{Figure}
	\centering
	\includegraphics[width=.9\textwidth]{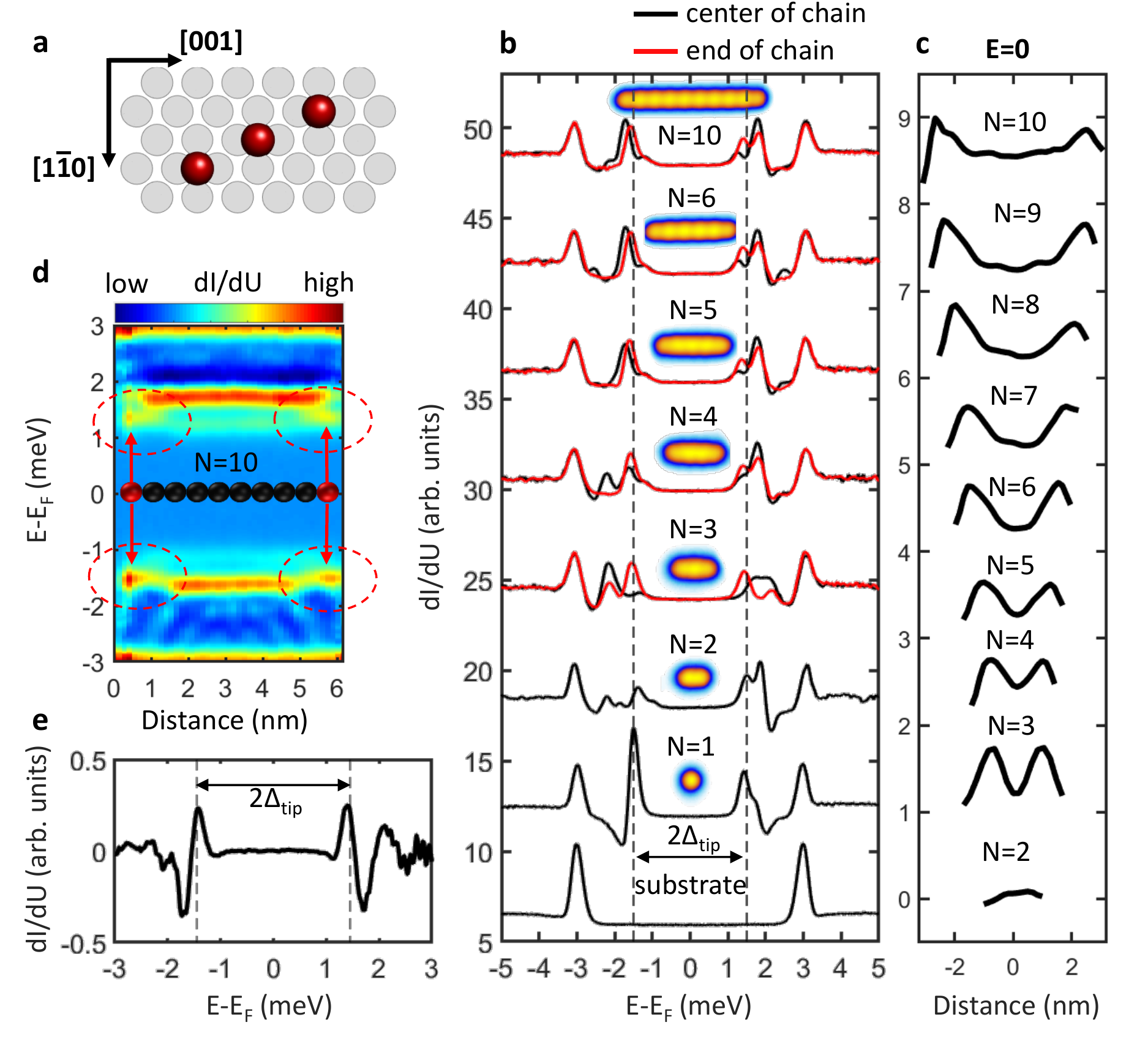}
	\caption{\textbf{Building a chain of Cr atoms, atom-by-atom. a} Illustration of the positions of the Cr atoms (red) on the Nb lattice. Atoms are added sequentially along the $[1\overline{1}3]$ direction at a spacing of \SI{0.55}{\nano\meter}; \textbf{b} dI/dU signal measured after sequentially adding Cr atoms to the chain, one atom at a time, up to a chain that is 10 atoms long. For chains longer than N= 6 up to 10 atoms, no significant changes in the energy dependence of dI/dU are observed. Black and red signals correspond to the center and end of the chains, respectively. \textbf{c} Spectral accumulation at zero energy (sum of intensities at $\pm\Delta_\mathrm{tip}$ in our case because of the use of superconducting tips) emerges at the chain ends starting from N=3. \textbf{d} Energy resolved dI/dU intensity for N=10 visualizing the appearance of end modes (see red arrows) located at $\pm\Delta_\mathrm{tip}$ and highlighted in \textbf{e} after subtracting the background inside the bulk of the chains.}
	\label{Figure1}
\end{figure*}

A rich spectroscopic scenario is visible by positioning the STM tip on top of a single Cr ad-atom (N=1)~\cite{KMG2021}. Several peaks emerge within the bare Nb substrate superconducting energy  gap. 
These peaks are direct fingerprints of magnetic impurity-superconductor interactions, with magnetic moments inducing  YSR quasi-particle resonances residing inside the superconducting energy gap. YSR states always appear in pairs that are energetically particle-hole symmetric with respect to the Fermi level~\cite{Yu.1965,Shiba.1968,Rusinov.1969,Yazdani.1997}. Their energy is directly linked to the strength of the exchange coupling with the superconducting condensate $J$ while the difference in electron- and hole-like intensity is related to the magnetic impurity being in a spin-screened or free-spin regime.  As demonstrated in Supplementary Note 2, mapping their spatial distribution allows one to clearly identify their orbital character~\cite{CRC2017,KMG2021}. In the present case, the d$_{z^2}$ orbital dominates near the Fermi level, corresponding to a deep YSR state \cite{PhysRevB.88.155420}. When the magnetic impurities are brought close to each other, the YSR states undergo a shift depending on the interaction between adatoms \cite{PhysRevB.67.020502,PhysRevB.73.224511},  with the individual YSR states hybridizing and creating the so-called Shiba bands.  %\st{Since the d$_{z^2}$-derived YSR states are very close to zero energy ($\pm\Delta_\mathrm{tip}$ in our case because of the use of superconducting tips), a small hybridization is already sufficient to drive a quantum phase transition where for electron- and hole- component of YSR pairs to cross through zero. This quantum phase transition (QPT) is a necessary but not sufficient  condition for the transition into a topologically non-trivial superconducting regime where zero-energy modes are trapped and localized at the end of the chain~\cite{PhysRevB.88.020407,PhysRevLett.111.186805,PhysRevB.88.155420}.}

Here we scrutinize in more detail the creation and evolution of Shiba bands  atom-by-atom  starting from the isolated atom case. Spectroscopic results are summarized in Fig.~\ref{Figure1}b.  When two adatoms are brought close to each other (N=2), their interaction shifts the d$_{z^2}$-derived YSR pairs towards higher binding energies as compared to the single adatom case (N=1) showing higher intensity above the Fermi level \cite{doi:10.1126/science.1202204,Dinge2024837118,kuster2021long}. When a third atom is added (N=3), spectral weight is trapped at zero energy at the end of the chain (see red curve). This becomes more evident by progressively increasing the length of the chain. The YSR bands in the bulk (center of the chain) do not show any significant spectral evolution, within the limit of our experimental resolution, for chains longer than N=6 atoms, signaling that the asymptotic limit is reached for a very short distance. The  spatial distribution of the zero-energy mode is reported in Fig.~\ref{Figure1}c, showing a localization of approximately \SI{1}{\nano\meter}, which is independent of the chain length. The localization at the chain ends is also visible in the full spectroscopy map acquired along the chain, reported in Fig.~\ref{Figure1}d. When normalized by subtracting the average spectrum in the bulk of the chain, the spectrum at the chain's end shows, within the limit of our energy resolution, particle-hole symmetric peaks centered at the tip superconducting energy gap $\pm\Delta_\mathrm{tip}$, as visualized in Fig.\ref{Figure1}e. As illustrated in Supplementary Figure 3, no significant change is observed on longer chains and by using different microtips. This signature would be compatible with one of the distinct fingerprints of Majorana modes. Moreover, the persistence of these modes at zero energy independently from the chain length allows  us to effectively identify them  as boundary modes.  However, it is highly unlikely that such short chains realize a topological superconductor, an observation which indicates that they are boundary states, plausibly of trivial origin, and accidentally located at zero energy. Indeed, it is conventionally expected that at such short distances Majorana states at opposite ends hybridize and split away from zero energy \cite{PhysRevB.87.094518}.

\subsection{Crystallographic direction and hopping dependence of boundary modes.}
\begin{figure*}
        \renewcommand{\figurename}{Figure}
	\centering
	\includegraphics[width=.95\textwidth]{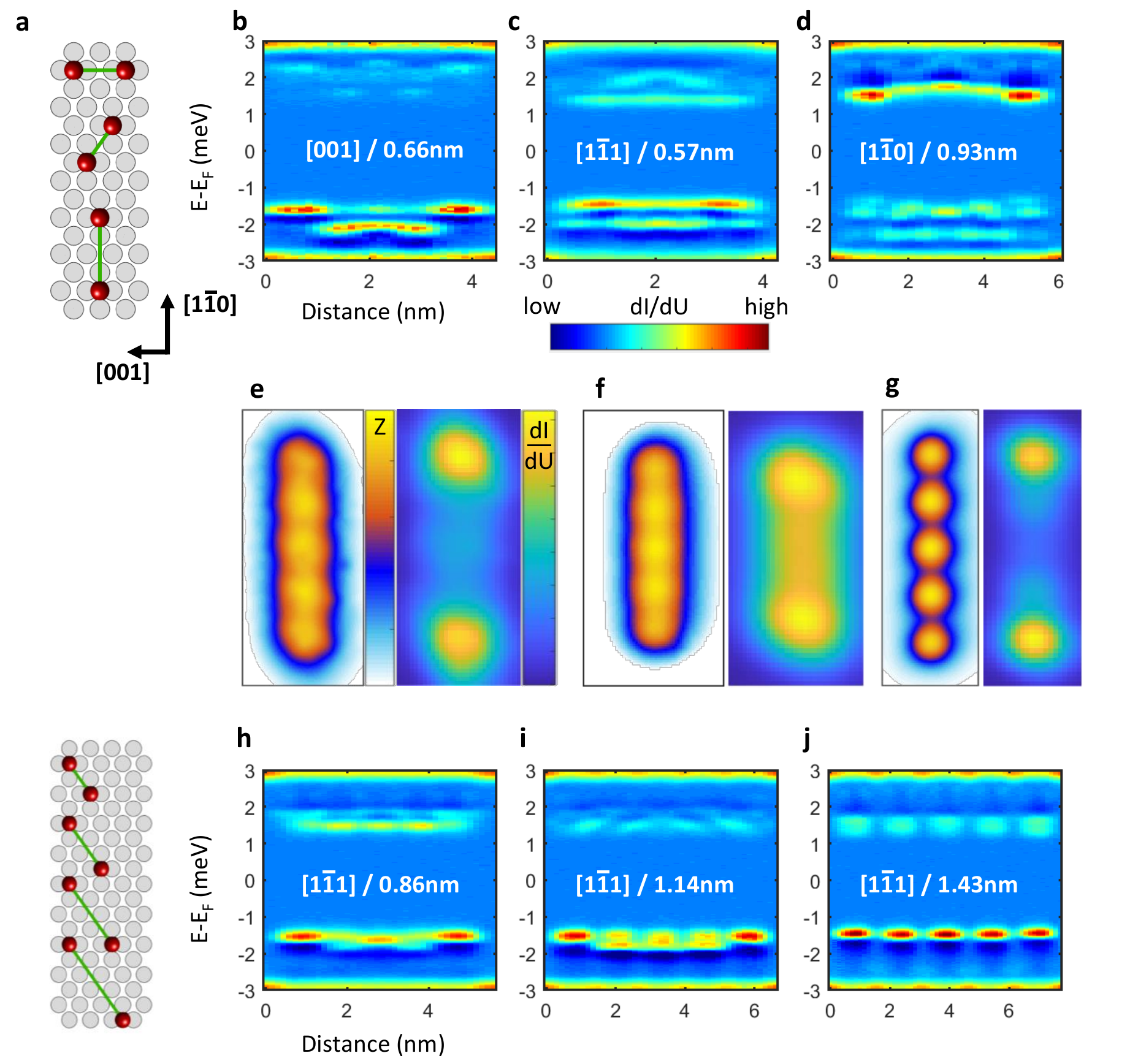}
	\caption{\textbf{Spin chains along different directions. a} Illustration of the Cr atom positions (red) on the Nb lattice for three different alignments (top panel) and four different spacing between adatoms (bottom panel). In each case a green line shows the connections between the atoms. \textbf{b-d} Spectroscopic mapping of the chain for each  of the scenarios illustrated in (a). \textbf{e-g} Accumulation of spectral weight at zero energy (sum of the intensities at $\pm\Delta_\mathrm{tip}$ in our case because of the superconducting tip energy gap) localized at the chains end.  All chains consist of five atoms. \textbf{h-j} Spectroscopic mapping of spin chains by progressively increasing the spacing between adatoms.}
	\label{Figure2}
\end{figure*}

The electronic structure anisotropy of the Nb(110) surface \cite{PhysRevB.102.174502} together with the presence of deep YSR pairs \cite{CRC2017,KMG2021} offers a unique opportunity to experimentally scrutinize different experimental scenarios. Indeed, spin chains built along various crystallographic directions are characterized  by distinct spacing among the atoms, resulting in different interaction strengths and magnetic ground states and, consequently, topological phase diagram (see section Discussion). Moreover, by varying the inter-atomic spacing inside the chain, it is possible to experimentally vary the hopping amplitude. Overall, this supplies unprecedented flexibility to controllably explore the formation of YSR bands in a large experimental parameter space. 
To shed light on these aspects, Figure \ref{Figure2} displays chains consisting of five atoms aligned along three distinct crystallographic directions, as schematically illustrated in Figure \ref{Figure2}a. The inter-atomic spacing is \SI{0.66}{\nano\meter}, \SI{0.57}{\nano\meter} and \SI{0.93}{\nano\meter} along  $[001]$, $[1\overline{1}1]$ and $[1\overline{1}0]$ directions, respectively.

Figure~\ref{Figure2}b-d shows the spectroscopic mapping of the superconducting state induced in each one of the chains. Its inspection reveals that, despite each direction being characterized by distinct spectroscopic signatures, all chains are characterized by a significant spectral weight accumulation at or very close to zero energy localized at the end already for very short chains, as visible in Figure~\ref{Figure2}e-g~\cite{PhysRevB.88.155420}. Note that, because of tip superconducting energy gap, the zero-energy spectral weight of the chains corresponds to the sum of the intensities at $\pm\Delta_\mathrm{tip}$. Additional spectroscopic data confirming the spectral weight accumulation for these as well as for longer chains are reported in Supplementary Notes 5 and 6.

This scenario can be directly controlled by acting on the hopping amplitude, which determines the bandwidth of the Shiba bands. For bandwidths smaller than the energy of the single atom YSR pairs, the resulting Shiba bands are well-separated, i.e. electron- and hole-components  avoid crossing each other at zero energy. This concept is illustrated in Fig.~\ref{Figure2}h-j for chains assembled along the $[1\overline{1}1]$ direction, i.e. the direction characterized by the smallest possible discrete distance between adatoms. These chains correspond to the experimental realization of the distinct configurations illustrated in the bottom panel of Fig.~\ref{Figure2}a. By increasing the distance between the adatoms (from h to j), the hopping amplitude is progressively reduced. Our results reveal that the strong accumulation of spectral weight predominantly localized at the end of the chains becomes very weak for distances of approximately \SI{1.45}{\nano\meter}.  Interestingly, a distinct boundary behavior at zero-energy is also found in the metallic regime~\cite{brinker2021anomalous} (see Supplementary Note 7 and discussion below).

\subsection{Dimerized chains.}

\begin{figure*}
        \renewcommand{\figurename}{Figure}
	\centering
	\includegraphics[width=\textwidth]{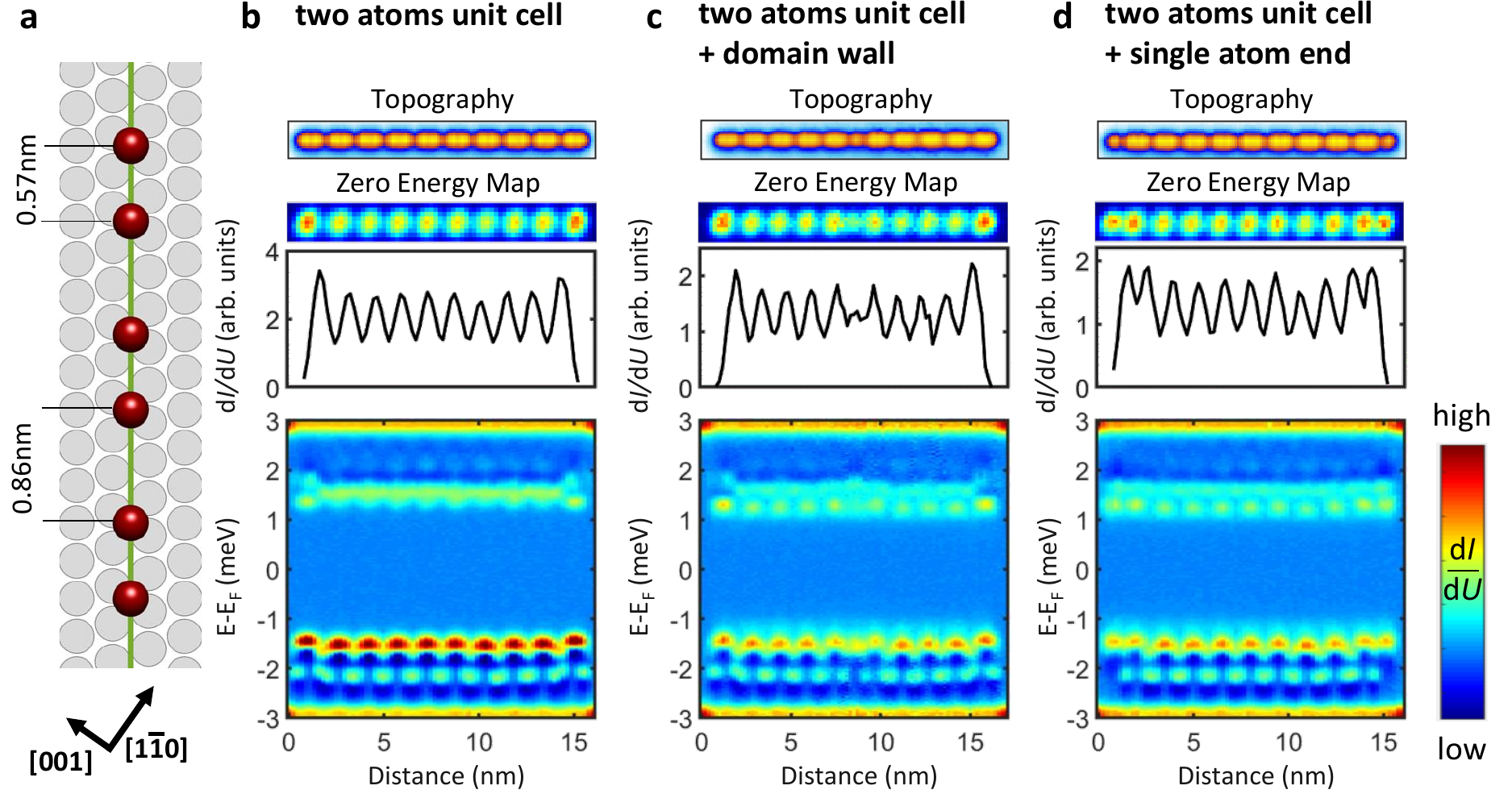}
	\caption{\textbf{Distinct unit cells and robustness against perturbations. a} Schematic illustration of a two-atoms unit cell spin chain. \textbf{b-d} Topography (top panels), zero energy spatial map (middle panels), and spectroscopic mapping  (bottom panels) for \textbf{b} the pristine chain, \textbf{c} chain with internal perturbation, \textbf{d} chain with perturbation at the end. Zero energy end modes (corresponding to the sum of the intensities at $\pm\Delta_\mathrm{tip}$ because of superconducting tip energy gap) are visible for all chains.}
	\label{Figure4}
\end{figure*}

To scrutinize the origin of the zero-energy end modes, we experimentally assembled chains where the unit cell consists of two adatoms. Despite its importance in clarifying the origin of the end modes, this scenario has never been explored so far, with all previous works focusing on chains consisting of equally spaced nearest neighboring adatoms \cite{Nadj-Perge602,PhysRevLett.115.197204,PKK2016,Kimeaar5251,KCW2018,SBS2020}. As schematically illustrated in Figure \ref{Figure4}a, a two-atom unit cell structure is achieved by periodically placing adatoms at two distinct distances corresponding to two distinct hopping amplitudes. Fig. \ref{Figure4}b shows a topographic image of the resulting spin chain, which consists of 10 unit cells assembled using 20 atoms. The spectroscopic measurements visualize a rather delocalized state with a stronger accumulation of spectral weight at zero-energy localized at the chain end, as highlighted by the zero energy map and its relative line profile (because of tip superconducting energy gap, the zero-energy spectral weight of the chains corresponds to the sum of the intensities at $\pm\Delta_\mathrm{tip}$). Its localization still takes place on a very short length scale (see zero energy maps in Figure \ref{Figure4}). To  clarify the origin of the zero-energy modes, we tested their robustness against perturbations. This is illustrated in Figure~\ref{Figure4}c-d. Figure~\ref{Figure4}c shows a chain where an additional adatom is placed inside the bulk that acts as a structural and electronic defect, acting as a domain wall. If the zero energy spectral intensity highly localized at the end of the chain would be a Majorana mode, we might expect to observe four: two at the chain ends, and two more on the right and on the left of the domain wall, because the domain wall would cut the topological superconductors into two, which is not in line with our observation.  However, a more complex scenario cannot be excluded. Indeed, the chain end and the domain wall are not fully equivalent cases. The defect provides a perturbation as well as a continuity in the electronic structure, which makes it distinct with respect to the chain end. In Figure~\ref{Figure4}d, additional single adatoms have been connected to the end of the chains. The spectral weight accumulation at zero energy remains localized at the chain end. However, a higher spectral intensity at zero is now observed on both the dimer and the last adatom. Since the bulk of the wire has not been modified, which dictates the topological properties and the intrinsic spatial extension of possible Majorana modes, their localization should not be altered by changing the edge.

%\subsection{Magnetism, topological invariant and origin of zero energy boundary modes.}

\subsection{Origin of zero-energy modes.}

\begin{figure*}[t!]
        \renewcommand{\figurename}{Figure}
	\centering
	\includegraphics[width=1\textwidth]{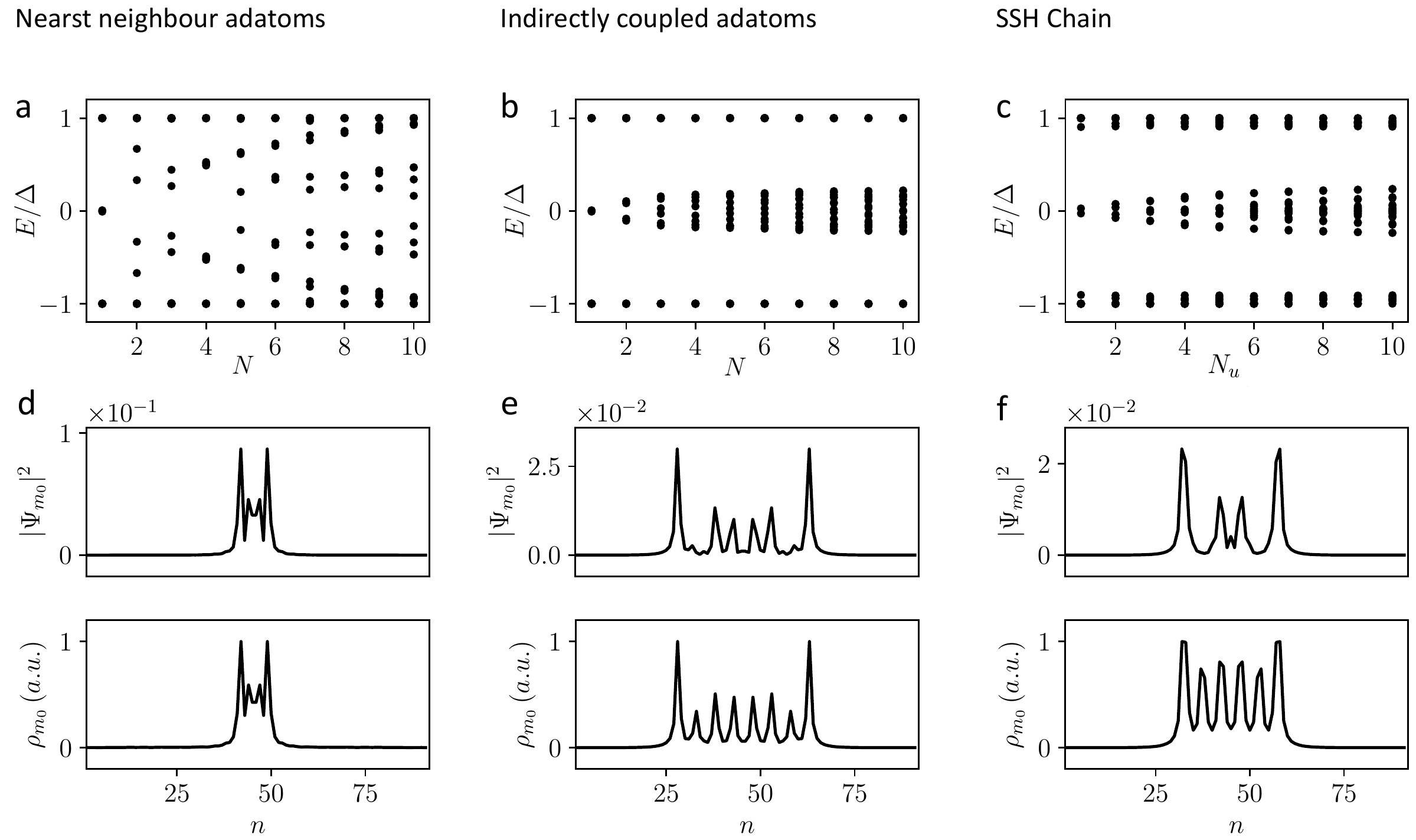}
	\caption{\textbf{Energies and wave functions for  different impurity chain configurations.}  The adatoms are separated by a distance  (\textbf{a, d}) $L=a$ and  (\textbf{b, e}) by $L=5a$.  (\textbf{c, f}) The impurities are ordered in unit cells: each unit cell contains two adatoms. We set the intra unit cell distance between the two adatoms to $L=a$ and the spatial separation between  the unit cells to $L=4a$.   Forty lowest energies  (\textbf{a, b}) as a function of the number of impurities $N$ and (\textbf{c}) as a function of the number of unit cells $N_u$. (\textbf{d-f}) Probability density and LDOS at zero energy at the cross-section of the chain. The parameters used for the figure are $t=1$, $\Delta=0.2$, $\mu=1$, $\epsilon=0.01$ and the total size is $N_x\times N_y = 50 \times 92$. Furthermore, we used  $J=J_c\approx 2.47$ in panels (\textbf{a, b, d, e}) and $J=J_<\approx 1.88$ in panels (\textbf{c, f}). }
	\label{FigureShibaStates}
\end{figure*}

In the following we explore theoretically whether the  investigated finite chains can host Majorana states by building up a minimal tight binding model (see details in Supplementary Note 8). In particular, we consider the trivial regime by neglecting the spin-orbit interaction (SOI). 
For a single magnetic impurity, the experimental data reveal a deep YSR state close to zero energy (see Supplementary Note 2). This occurs for a certain $s-d$ exchange coupling strength $J = J_c$, between the adatom's spin and that of the surface electrons, which we fix in our model, unless stated otherwise. 
Adding further nearby magnetic impurities leads to hybridization and energy splitting of the YSR states as a function of their overlap, which is determined by the distance $L$ between the impurities. If the impurities are deposited close to each other, e.g. only separated by one lattice spacing, the energy of the lowest sub-gap state strongly oscillates as  a function  of  the  chain length,  as shown in Fig.~\ref{FigureShibaStates}a. For specific chain lengths, the lowest energy state can be experimentally detected as a zero-energy mode  if the finite energy resolution is taken into account. Moreover, the probability density of the lowest sub-gap state and the local density of states (LDOS, which accounts also for higher states due to the finite-energy broadening $\epsilon$) are both characterized by large weights on the ends of the chain, as illustrated in Fig.~\ref{FigureShibaStates}d.

Next, we consider well-separated impurities at a distance $L=5a$ to simulate the experimental scenario scrutinized in the present study, i.e. the indirectly coupled regime. In this case, the splitting is much weaker and the resulting YSR band appears narrow and localized around zero energy, as shown in Fig.~\ref{FigureShibaStates}b. In other words, the system hosts states with almost zero energy independent of the length of the chain. For some parameter choices the probability density of the lowest sub-gap state along the chain reveals large weights even on the ends of the chains, see Fig.~\ref{FigureShibaStates}e, matching the experimental observation (see Fig.~\ref{Figure1} and Fig.~\ref{Figure2}). Since the YSR states in the model are energetically close to each other, we calculate the LDOS with a broadening parameter $\epsilon$ to account for the finite experimental energy resolution. The LDOS reveals, similar to the probability density, large peaks at the ends of the chain (see Fig.~\ref{FigureShibaStates}e), effectively reproducing the experimental observations. Therefore, without claiming that the experimental situation is exactly captured by the minimal tight binding model, we conclude that trivial mechanisms can lead to effects similar as those observed in the experiments.
 
Finally, we order the impurities in unit cells. In particular, two impurities are separated by one lattice spacing $a$ in each unit cell and the unit cells are separated  by a distance of $L=4a$.  In order to find trivial  zero-energy states, we start by calculating the energies of a system consisting of one unit cell as a function of the exchange coupling strength. The two  overlapping YSR states have zero energy at $J_<$ and $J_>$, respectively, with  $J_<<J_c<J_>$.  Finally, we set the exchange coupling strength to $J_<$ and calculate the energies as a function of the number of unit cells $N_u$. The unit cells are well separated and a narrow band forms around zero energy, as shown in Fig.~\ref{FigureShibaStates}c. The probability density of the lowest state has again strong weights on the ends of the chain  and the LDOS (see Fig.~\ref{FigureShibaStates}f) looks similar to the experimental $dI/dV$ zero-energy map reported in Fig.~\ref{Figure4}b. We would like to note that the effect of having a higher intensity at the end is not generally valid for the probability density of the lowest state, i.e. for other chain lengths the weight can be distributed differently. However the LDOS, which takes multiple states close to zero energy into account, has  almost the same shape also for varying chain length, namely peaks at the positions of the unit cells at slightly higher peaks at the ends of the chain.

\section*{Discussion}

\begin{figure*}%[!h]
	\renewcommand{\figurename}{Figure}
	\centering
	\includegraphics[width=.85\textwidth]{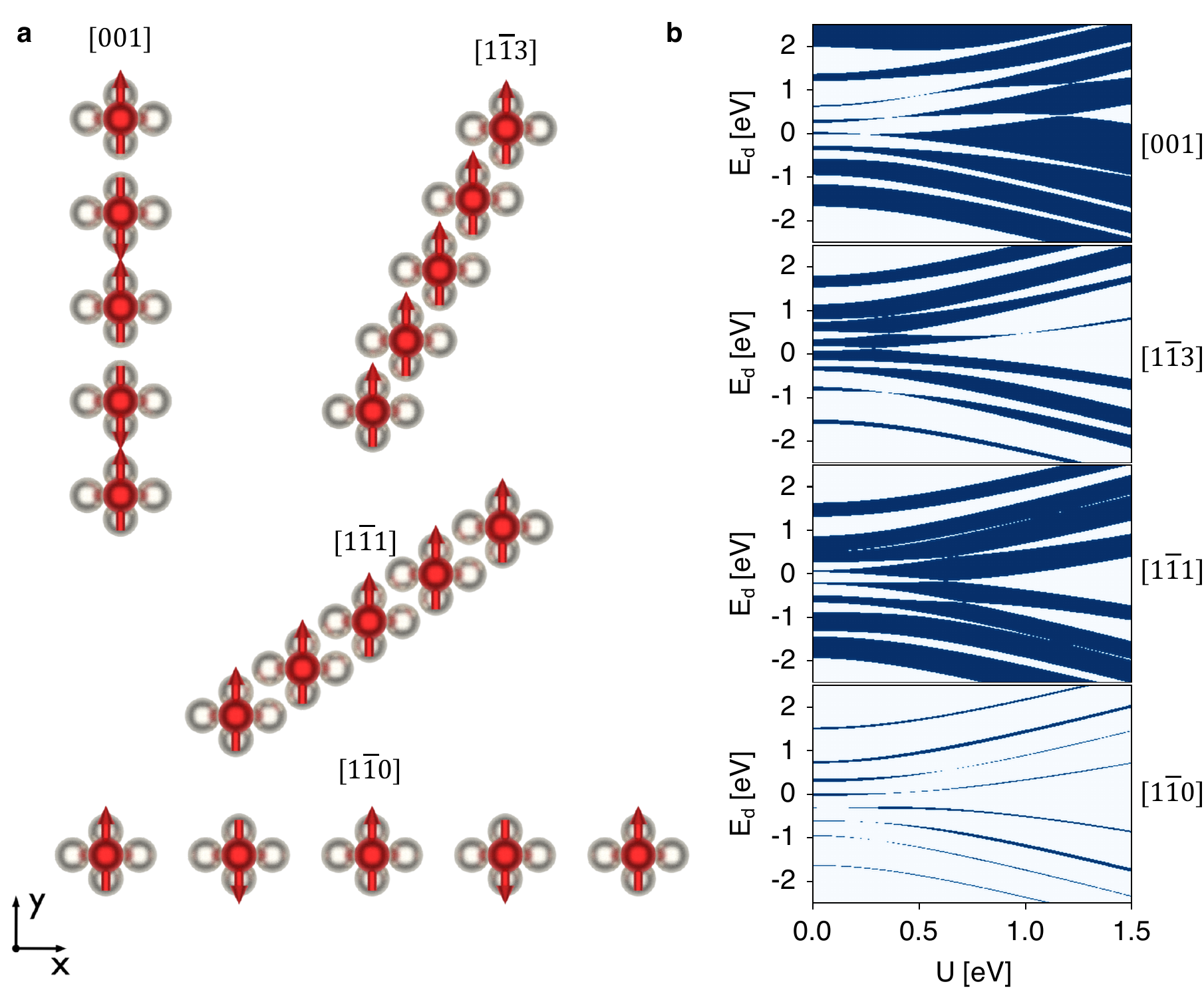}
	\caption{
		\textbf{Topological properties and magnetic ground states of  the different Cr chains deposited on the Nb(110) surface.} Shown are the $[001]$, $[1\overline{1}3]$, $[1\overline{1}1]$, and $[1\overline{1}0]$ directions.
		\textbf{a} Magnetic structures obtained from first principles (see Methods). 
		\textbf{b} Phase diagrams of the Majorana number for the infinite wires as function of the spin splitting $U$ and the energy of the $d$-orbitals $E_d$ and including hopping up to four nearest neighbors. The range of hopping can induce destructive or constructive interference effects. For example, the non-trivial regions are amplified by considering four neighbors instead of only nearest neighbors for the $[001]$ chain, while they are reduced in size for the $[1\overline{1}0]$ chain.
		Topological regions are shown in blue, while the trivial non-topological regions are white. 
		For more details see Supplementary Notes 9 and 10.
	}
	\label{Theory_Figure}
\end{figure*}
Our results systematically scrutinize dilute spin chains proximitized to a superconducting condensate. Contrary to earlier studies on ferromagnetically coupled and densely packed adatoms \cite{Nadj-Perge602,PhysRevLett.115.197204,PKK2016,Kimeaar5251,KCW2018,SBS2020,PhysRevB.104.045406}, our platform effectively uses the superconducting condensate as an active element to mediate the interaction between adatoms. Starting from single adatoms hosting deep Shiba states, we demonstrate the ability to create distinct magnetic ground state configurations hosting Shiba bands crossing the Fermi level. This is associated to the emergence of  highly localized boundary modes showing some spectroscopic signatures in agreement with expectations for Majorana modes. However, these are identified as trivial Shiba states characterized by a larger spectral weight localized at the chain ends. 

Even if in theory the chains are made long enough (beyond our current experimental reach), a map of the topological invariants for infinite Cr wires along distinct crystallographic directions and characterized by different magnetic states (see Figure~\ref{Theory_Figure}a, Method section and Supplementary Notes 9 for more details) clearly indicates the sensitivity of the topological behavior on the details of the electronic structure. This is illustrated in Figure~\ref{Theory_Figure}b, which summarizes the results of multi-orbital tight-binding simulations parametrized from ab-initio and including the proximity-induced superconductivity as a parameter (see Supplementary Note 10). Very small topologically non-trivial regions (blue areas) appear for the $[1\overline{1}0]$ chain, which is characterized by weak hopping and flat electronic bands. Increasing the hopping amplitude enlarges the bandwidths and the size of the topological non-trivial regions, as can be seen in the $[1\overline{1}1]$, $[1\overline{1}3]$ and $[001]$ chains. For the latter, the hopping is comparable to the crystal field splitting, leading to an overlap of the different bands and a complex phase diagram. This shows that, even if the wires are made long enough, entering into a topologically non-trivial regime is challenging once considering the complexity of real materials. It requires fine control of different parameters such as exchange splitting, position of d-orbital, range of hopping, and crystal field splitting.

Overall, our results imply that zero-bias, length independent, boundary modes are not necessarily linked to Majorana modes. At the same time, by using indirectly coupling mechanisms on anisotropic surfaces, we demonstrate that it is possible to significantly enlarge the experimental parameter space, opening new routes for the experimental search of topologically non-trivial superconducting states.

{\it{Note.}} After completing our work, we became aware of a STM-based study on diluted chains of Fe atoms coupled to superconducting NbSe$_2$~\cite{liebhaber2021quantum}. 

%\section*{Methods}
\begin{methods}

\subsection{Sample and tip preparation.}
A clean Nb(110) substrate was obtained by flashing the crystal to \SI{2300}{\kelvin} in \SI{12}{\second}~\cite{Odobesko.2019} hundreds of times. Measurements were taken at a temperature T = \SI{600}{\milli\kelvin} using the Tribus STM head (Scienta Omicron). Cr adatoms were deposited in-situ using an e-beam evaporator while keping the sample below a temperature of \SI{15}{\kelvin}. All atoms were found to be adsorbed in the hollow site of the Nb(110) surface~\cite{Kuester2021}.  Cr chains have been assembled by atomic manipulation technique by dragging them with the STM tip in constant current mode using a setpoint of V = \SI{-5}{\milli\volt}; I = \SI{70}{\nano\ampere}. $\mathrm{d}I/\mathrm{d}U$ spectra were acquired by standard lock-in technique. To enhance the energy resolution, we used superconducting Nb tips obtained by deep indentations into the Nb single crystal. 

\subsection{First-principles and tight-binding model calculations}
%\sascha{mainly copied from dimer paper and adapted}
The ab-initio simulations are based on the scalar-relativistic full-electron Korringa-Kohn-Rostoker (KKR) Green function augmented self-consistently with spin-orbit interaction~\cite{Papanikolaou:2002,Bauer:2014}.
The theoretical framework uses multiple-scattering theory allowing an embedding scheme, which is ideal to address the electronic and magnetic properties of nanostructures in real space without the use of periodic supercells. We utilize the atomic-sphere approximation (ASA) and consider the full charge density within the local spin density approximation (LSDA)~\cite{Vosko:1980}.
We assume an angular momentum cutoff at $\ell_{\text{max}} = 3$ for the orbital expansion of the Green function and when extracting the local density of states a k-mesh of $150 \times 150$ is considered.
The Nb(110) surface is modelled by slab containing 22 layers enclosed by two vacuum regions with a thickness of \SI{9.33}{\angstrom} each.
Due to the large spacings between the atoms of the Cr chains we use the geometrical properties obtained for isolated Cr adatoms.
Thus, the chain atoms are placed on the hollow stacking site relaxed towards the surface by \SI{20}{\percent} of the inter-layer distance of the underlying Nb(110) surface, which was shown to be the energetically favoured stacking for isolated adatoms in Ref.~\cite{Kuester2021}.\\

Due to the large distances between the Cr atoms in all chains, we find negligible  boundary effects in the electronic and magnetic ground-state structures of the chains.
The magnetic moments of approximately $\SI{3.0}{\mub}$ per atom are uniformly distributed with a difference of less than \SI{0.02}{\mub} between the edge and the central atoms for all chains.

To describe the magnetic structure of the chains we use the generalized Heisenberg model, 
\begin{equation}
    H = \sum_i \vec{e}_i \mathcal{K}_i \vec{e}_i + \frac{1}{2} \sum_{ij} J_{ij} \vec{e}_i \cdot \vec{e}_j + \frac{1}{2} \sum_{ij} \vec{D}_{ij} \cdot \left( \vec{e}_i \times \vec{e}_j \right) \quad ,
\end{equation}
where $\mathcal{K}_i$ is the on-site anisotropy, $J_{ij}$ is the isotropic exchange interaction and $\vec{D}_{ij}$ is the Dzyalo\-shinskii-Moriya interaction between the $i$-th and $j$-th atom with the magnetic moment $\vec{m}_i = m_i \vec{e}_i$.
The magnetic exchange interactions were obtained using the magnetic force theorem in the frozen-potential approximation and the infinitesimal rotation method \cite{Liechtenstein1987,Ebert2009}.
The on-site magnetic anisotropy of the Cr atoms in all chains is obtained from the method of constraining fields \cite{Brinker2019}, and was found to be similar to the one of the isolated adatoms reported in Ref.~\cite{Kuester2021}.
Further information can be found in Supplementary Note 4.
\\
The different chains exhibit fundamentally different magnetic exchange interactions among nearest neighboring adatoms, ranging from antiferromagnetic coupling of $J=\SI{2.1}{\milli\electronvolt}$ for the $[1\overline{1}0]$ chain and $J=\SI{11. 5}{\milli\electronvolt}$ for the $[001]$ chain to ferromagnetic coupling of $J=\SI{-6.2}{\milli\electronvolt}$ for the $[1\overline{1}1]$ chain and $J=\SI{-1.7}{\milli\electronvolt}$ for the $[1\overline{1}3]$ chain.  
Minimizing a Heisenberg model with the magnetic exchange parameters from first principles leads to the ground states of the chains shown in Figure \ref{Theory_Figure}a.
Due to the weak Dzyaloshinskii-Moriya interaction, the chains are almost collinear and follow the ferro- and antiferromagnetic couplings dictated by the isotropic exchange (see Methods section and Supplementary Note 6 for more details).
\\
The topological invariant is calculated using a multi-orbital  tight-binding model with parameters obtained from density functional theory and an effective Hamiltonian construction (see Ref.~\cite{SBS2020} and Supplementary Note 7).

\end{methods}

\begin{addendum}

%\item[Authors contributions]  S.L. and P.S. conceived the project. F.K. performed the STM measurements and 
%analysed the experimental data under supervision of S.S.P. and P.S.. S.B. performed the ab-initio simulations. S.B. and S.L. conceived the mapping approach from first-principles to tight-binding and studied the related computed data. R.H, D.L. and J.K developed the minimal tight-binding model, performed simulations and analysed the associated results. P.S. wrote the initial version of the manuscript to which all authors contributed.

%\item[Competing Interests] The authors declare that they have no competing interests. 

\item[Data and materials  availability] All data needed to evaluate the conclusions in the paper are present in the paper and/or the supplementary materials. Additional data related to this paper may be requested from the authors.  The KKR Green function code that supports the findings of this study is available from the corresponding author on reasonable request.
\end{addendum}

\section*{References}
\bibliographystyle{naturemag}
\bibliography{bibliography}

\begin{thebibliography}{10}
\expandafter\ifx\csname url\endcsname\relax
  \def\url#1{\texttt{#1}}\fi
\expandafter\ifx\csname urlprefix\endcsname\relax\def\urlprefix{URL }\fi
\providecommand{\bibinfo}[2]{#2}
\providecommand{\eprint}[2][]{\url{#2}}

\bibitem{Alicea_2012}
\bibinfo{author}{Alicea, J.}
\newblock \bibinfo{title}{New directions in the pursuit of majorana fermions in
  solid state systems}.
\newblock \emph{\bibinfo{journal}{Reports on Progress in Physics}}
  \textbf{\bibinfo{volume}{75}}, \bibinfo{pages}{076501}
  (\bibinfo{year}{2012}).
\newblock \urlprefix\url{https://doi.org/10.1088/0034-4885/75/7/076501}.

\bibitem{Beenakker}
\bibinfo{author}{Beenakker, C.}
\newblock \bibinfo{title}{Search for majorana fermions in superconductors}.
\newblock \emph{\bibinfo{journal}{Annual Review of Condensed Matter Physics}}
  \textbf{\bibinfo{volume}{4}}, \bibinfo{pages}{113--136}
  (\bibinfo{year}{2013}).
\newblock
  \urlprefix\url{https://doi.org/10.1146/annurev-conmatphys-030212-184337}.
\newblock \eprint{https://doi.org/10.1146/annurev-conmatphys-030212-184337}.

\bibitem{Sato_2017}
\bibinfo{author}{Sato, M.} \& \bibinfo{author}{Ando, Y.}
\newblock \bibinfo{title}{Topological superconductors: a review}.
\newblock \emph{\bibinfo{journal}{Reports on Progress in Physics}}
  \textbf{\bibinfo{volume}{80}}, \bibinfo{pages}{076501}
  (\bibinfo{year}{2017}).
\newblock \urlprefix\url{https://doi.org/10.1088/1361-6633/aa6ac7}.

\bibitem{Mourik1003}
\bibinfo{author}{Mourik, V.} \emph{et~al.}
\newblock \bibinfo{title}{Signatures of majorana fermions in hybrid
  superconductor-semiconductor nanowire devices}.
\newblock \emph{\bibinfo{journal}{Science}} \textbf{\bibinfo{volume}{336}},
  \bibinfo{pages}{1003--1007} (\bibinfo{year}{2012}).
\newblock \urlprefix\url{https://science.sciencemag.org/content/336/6084/1003}.
\newblock
  \eprint{https://science.sciencemag.org/content/336/6084/1003.full.pdf}.

\bibitem{Nadj-Perge602}
\bibinfo{author}{Nadj-Perge, S.} \emph{et~al.}
\newblock \bibinfo{title}{Observation of majorana fermions in ferromagnetic
  atomic chains on a superconductor}.
\newblock \emph{\bibinfo{journal}{Science}} \textbf{\bibinfo{volume}{346}},
  \bibinfo{pages}{602--607} (\bibinfo{year}{2014}).
\newblock \urlprefix\url{https://science.sciencemag.org/content/346/6209/602}.
\newblock
  \eprint{https://science.sciencemag.org/content/346/6209/602.full.pdf}.

\bibitem{PhysRevLett.115.197204}
\bibinfo{author}{Ruby, M.} \emph{et~al.}
\newblock \bibinfo{title}{End states and subgap structure in proximity-coupled
  chains of magnetic adatoms}.
\newblock \emph{\bibinfo{journal}{Phys. Rev. Lett.}}
  \textbf{\bibinfo{volume}{115}}, \bibinfo{pages}{197204}
  (\bibinfo{year}{2015}).
\newblock
  \urlprefix\url{https://link.aps.org/doi/10.1103/PhysRevLett.115.197204}.

\bibitem{PKK2016}
\bibinfo{author}{Pawlak, R.} \emph{et~al.}
\newblock \bibinfo{title}{Probing atomic structure and majorana wavefunctions
  in mono-atomic fe chains on superconducting pb surface}.
\newblock \emph{\bibinfo{journal}{npj Quantum Information}}
  \textbf{\bibinfo{volume}{2}}, \bibinfo{pages}{16035} (\bibinfo{year}{2016}).
\newblock \urlprefix\url{https://doi.org/10.1038/npjqi.2016.35}.

\bibitem{Kimeaar5251}
\bibinfo{author}{Kim, H.} \emph{et~al.}
\newblock \bibinfo{title}{Toward tailoring majorana bound states in
  artificially constructed magnetic atom chains on elemental superconductors}.
\newblock \emph{\bibinfo{journal}{Science Advances}}
  \textbf{\bibinfo{volume}{4}} (\bibinfo{year}{2018}).
\newblock \urlprefix\url{https://advances.sciencemag.org/content/4/5/eaar5251}.
\newblock
  \eprint{https://advances.sciencemag.org/content/4/5/eaar5251.full.pdf}.

\bibitem{KCW2018}
\bibinfo{author}{Kamlapure, A.}, \bibinfo{author}{Cornils, L.},
  \bibinfo{author}{Wiebe, J.} \& \bibinfo{author}{Wiesendanger, R.}
\newblock \bibinfo{title}{Engineering the spin couplings in atomically crafted
  spin chains on an elemental superconductor}.
\newblock \emph{\bibinfo{journal}{Nature Communications}}
  \textbf{\bibinfo{volume}{9}}, \bibinfo{pages}{3253} (\bibinfo{year}{2018}).
\newblock \urlprefix\url{https://doi.org/10.1038/s41467-018-05701-8}.

\bibitem{PhysRevLett.114.017001}
\bibinfo{author}{Xu, J.-P.} \emph{et~al.}
\newblock \bibinfo{title}{Experimental detection of a majorana mode in the core
  of a magnetic vortex inside a topological insulator-superconductor
  ${\mathrm{bi}}_{2}{\mathrm{te}}_{3}/{\mathrm{nbse}}_{2}$ heterostructure}.
\newblock \emph{\bibinfo{journal}{Phys. Rev. Lett.}}
  \textbf{\bibinfo{volume}{114}}, \bibinfo{pages}{017001}
  (\bibinfo{year}{2015}).
\newblock
  \urlprefix\url{https://link.aps.org/doi/10.1103/PhysRevLett.114.017001}.

\bibitem{Majorana}
\bibinfo{author}{Majorana, E.}
\newblock \bibinfo{title}{Teoria simmetrica dell'elettrone e del positrone}.
\newblock \emph{\bibinfo{journal}{Il Nuovo Cimento (1924-1942)}}
  \textbf{\bibinfo{volume}{14}}, \bibinfo{pages}{171} (\bibinfo{year}{2008}).
\newblock \urlprefix\url{https://doi.org/10.1007/BF02961314}.

\bibitem{Wilczek2009}
\bibinfo{author}{Wilczek, F.}
\newblock \bibinfo{title}{Majorana returns}.
\newblock \emph{\bibinfo{journal}{Nature Physics}}
  \textbf{\bibinfo{volume}{5}}, \bibinfo{pages}{614--618}
  (\bibinfo{year}{2009}).
\newblock \urlprefix\url{https://doi.org/10.1038/nphys1380}.

\bibitem{AOR2011}
\bibinfo{author}{Alicea, J.}, \bibinfo{author}{Oreg, Y.},
  \bibinfo{author}{Refael, G.}, \bibinfo{author}{von Oppen, F.} \&
  \bibinfo{author}{Fisher, M. P.~A.}
\newblock \bibinfo{title}{Non-abelian statistics and topological quantum
  information processing in 1d wire networks}.
\newblock \emph{\bibinfo{journal}{Nature Physics}}
  \textbf{\bibinfo{volume}{7}}, \bibinfo{pages}{412--417}
  (\bibinfo{year}{2011}).
\newblock \urlprefix\url{https://doi.org/10.1038/nphys1915}.

\bibitem{Kitaev_2001}
\bibinfo{author}{Kitaev, A.~Y.}
\newblock \bibinfo{title}{Unpaired majorana fermions in quantum wires}.
\newblock \emph{\bibinfo{journal}{Physics-Uspekhi}}
  \textbf{\bibinfo{volume}{44}}, \bibinfo{pages}{131--136}
  (\bibinfo{year}{2001}).
\newblock \urlprefix\url{https://doi.org/10.1070/1063-7869/44/10s/s29}.

\bibitem{PhysRevLett.100.096407}
\bibinfo{author}{Fu, L.} \& \bibinfo{author}{Kane, C.~L.}
\newblock \bibinfo{title}{Superconducting proximity effect and majorana
  fermions at the surface of a topological insulator}.
\newblock \emph{\bibinfo{journal}{Phys. Rev. Lett.}}
  \textbf{\bibinfo{volume}{100}}, \bibinfo{pages}{096407}
  (\bibinfo{year}{2008}).
\newblock
  \urlprefix\url{https://link.aps.org/doi/10.1103/PhysRevLett.100.096407}.

\bibitem{PhysRevLett.105.077001}
\bibinfo{author}{Lutchyn, R.~M.}, \bibinfo{author}{Sau, J.~D.} \&
  \bibinfo{author}{Das~Sarma, S.}
\newblock \bibinfo{title}{Majorana fermions and a topological phase transition
  in semiconductor-superconductor heterostructures}.
\newblock \emph{\bibinfo{journal}{Phys. Rev. Lett.}}
  \textbf{\bibinfo{volume}{105}}, \bibinfo{pages}{077001}
  (\bibinfo{year}{2010}).
\newblock
  \urlprefix\url{https://link.aps.org/doi/10.1103/PhysRevLett.105.077001}.

\bibitem{PhysRevLett.105.177002}
\bibinfo{author}{Oreg, Y.}, \bibinfo{author}{Refael, G.} \&
  \bibinfo{author}{von Oppen, F.}
\newblock \bibinfo{title}{Helical liquids and majorana bound states in quantum
  wires}.
\newblock \emph{\bibinfo{journal}{Phys. Rev. Lett.}}
  \textbf{\bibinfo{volume}{105}}, \bibinfo{pages}{177002}
  (\bibinfo{year}{2010}).
\newblock
  \urlprefix\url{https://link.aps.org/doi/10.1103/PhysRevLett.105.177002}.

\bibitem{PhysRevB.88.020407}
\bibinfo{author}{Nadj-Perge, S.}, \bibinfo{author}{Drozdov, I.~K.},
  \bibinfo{author}{Bernevig, B.~A.} \& \bibinfo{author}{Yazdani, A.}
\newblock \bibinfo{title}{Proposal for realizing majorana fermions in chains of
  magnetic atoms on a superconductor}.
\newblock \emph{\bibinfo{journal}{Phys. Rev. B}} \textbf{\bibinfo{volume}{88}},
  \bibinfo{pages}{020407} (\bibinfo{year}{2013}).
\newblock \urlprefix\url{https://link.aps.org/doi/10.1103/PhysRevB.88.020407}.

\bibitem{PhysRevLett.111.186805}
\bibinfo{author}{Klinovaja, J.}, \bibinfo{author}{Stano, P.},
  \bibinfo{author}{Yazdani, A.} \& \bibinfo{author}{Loss, D.}
\newblock \bibinfo{title}{Topological superconductivity and majorana fermions
  in rkky systems}.
\newblock \emph{\bibinfo{journal}{Phys. Rev. Lett.}}
  \textbf{\bibinfo{volume}{111}}, \bibinfo{pages}{186805}
  (\bibinfo{year}{2013}).
\newblock
  \urlprefix\url{https://link.aps.org/doi/10.1103/PhysRevLett.111.186805}.

\bibitem{PhysRevB.88.155420}
\bibinfo{author}{Pientka, F.}, \bibinfo{author}{Glazman, L.~I.} \&
  \bibinfo{author}{von Oppen, F.}
\newblock \bibinfo{title}{Topological superconducting phase in helical shiba
  chains}.
\newblock \emph{\bibinfo{journal}{Phys. Rev. B}} \textbf{\bibinfo{volume}{88}},
  \bibinfo{pages}{155420} (\bibinfo{year}{2013}).
\newblock \urlprefix\url{https://link.aps.org/doi/10.1103/PhysRevB.88.155420}.

\bibitem{KHV2020}
\bibinfo{author}{Kezilebieke, S.} \emph{et~al.}
\newblock \bibinfo{title}{Topological superconductivity in a van der waals
  heterostructure}.
\newblock \emph{\bibinfo{journal}{Nature}} \textbf{\bibinfo{volume}{588}},
  \bibinfo{pages}{424--428} (\bibinfo{year}{2020}).
\newblock \urlprefix\url{https://doi.org/10.1038/s41586-020-2989-y}.

\bibitem{SBS2020}
\bibinfo{author}{Schneider, L.} \emph{et~al.}
\newblock \bibinfo{title}{Controlling in-gap end states by linking nonmagnetic
  atoms and artificially-constructed spin chains on superconductors}.
\newblock \emph{\bibinfo{journal}{Nature Communications}}
  \textbf{\bibinfo{volume}{11}}, \bibinfo{pages}{4707} (\bibinfo{year}{2020}).
\newblock \urlprefix\url{https://doi.org/10.1038/s41467-020-18540-3}.

\bibitem{SBP2021}
\bibinfo{author}{Schneider, L.} \emph{et~al.}
\newblock \bibinfo{title}{Topological shiba bands in artificial spin chains on
  superconductors}.
\newblock \emph{\bibinfo{journal}{Nature Physics}}  (\bibinfo{year}{2021}).
\newblock \urlprefix\url{https://doi.org/10.1038/s41567-021-01234-y}.

\bibitem{PhysRevB.104.045406}
\bibinfo{author}{Mier, C.} \emph{et~al.}
\newblock \bibinfo{title}{Atomic manipulation of in-gap states in the
  $\ensuremath{\beta}\ensuremath{-}{\mathrm{bi}}_{2}\mathrm{Pd}$
  superconductor}.
\newblock \emph{\bibinfo{journal}{Phys. Rev. B}}
  \textbf{\bibinfo{volume}{104}}, \bibinfo{pages}{045406}
  (\bibinfo{year}{2021}).
\newblock \urlprefix\url{https://link.aps.org/doi/10.1103/PhysRevB.104.045406}.

\bibitem{Dinge2024837118}
\bibinfo{author}{Ding, H.} \emph{et~al.}
\newblock \bibinfo{title}{Tuning interactions between spins in a
  superconductor}.
\newblock \emph{\bibinfo{journal}{Proceedings of the National Academy of
  Sciences}} \textbf{\bibinfo{volume}{118}} (\bibinfo{year}{2021}).
\newblock \urlprefix\url{https://www.pnas.org/content/118/14/e2024837118}.
\newblock \eprint{https://www.pnas.org/content/118/14/e2024837118.full.pdf}.

\bibitem{kuster2021long}
\bibinfo{author}{Küster, F.}, \bibinfo{author}{Brinker, S.},
  \bibinfo{author}{Lounis, S.}, \bibinfo{author}{Parkin, S. S.~P.} \&
  \bibinfo{author}{Sessi, P.}
\newblock \bibinfo{title}{Long range and highly tunable coupling between local
  spins coupled to a superconducting condensate} (\bibinfo{year}{2021}).
\newblock \eprint{2106.14932}.

\bibitem{KMG2021}
\bibinfo{author}{K{\"u}ster, F.} \emph{et~al.}
\newblock \bibinfo{title}{Correlating josephson supercurrents and shiba states
  in quantum spins unconventionally coupled to superconductors}.
\newblock \emph{\bibinfo{journal}{Nature Communications}}
  \textbf{\bibinfo{volume}{12}}, \bibinfo{pages}{1108} (\bibinfo{year}{2021}).
\newblock \urlprefix\url{https://doi.org/10.1038/s41467-021-21347-5}.

\bibitem{Yu.1965}
\bibinfo{author}{Yu, L.}
\newblock \bibinfo{title}{Bound state in superconductors with paramagnetic
  impurities}.
\newblock \emph{\bibinfo{journal}{Acta Physica Sinica}}
  \textbf{\bibinfo{volume}{21}}, \bibinfo{pages}{75--98}
  (\bibinfo{year}{1965}).
\newblock \urlprefix\url{http://wulixb.iphy.ac.cn/en/article/id/851}.

\bibitem{Shiba.1968}
\bibinfo{author}{Shiba, H.}
\newblock \bibinfo{title}{Classical spins in superconductors}.
\newblock \emph{\bibinfo{journal}{Progress of Theoretical Physics}}
  \textbf{\bibinfo{volume}{40}}, \bibinfo{pages}{435--451}
  (\bibinfo{year}{1968}).

\bibitem{Rusinov.1969}
\bibinfo{author}{Rusinov, A.~I.}
\newblock \bibinfo{title}{On the theory of gapless superconductivity in alloys
  containing paramagnetic impurities}.
\newblock \emph{\bibinfo{journal}{JETP}} \textbf{\bibinfo{volume}{29}},
  \bibinfo{pages}{1101--1106} (\bibinfo{year}{1969}).
\newblock \urlprefix\url{http://www.jetp.ac.ru/cgi-bin/dn/e_029_06_1101.pdf}.

\bibitem{Yazdani.1997}
\bibinfo{author}{Yazdani}, \bibinfo{author}{Jones}, \bibinfo{author}{Lutz},
  \bibinfo{author}{Crommie} \& \bibinfo{author}{Eigler}.
\newblock \bibinfo{title}{Probing the local effects of magnetic impurities on
  superconductivity}.
\newblock \emph{\bibinfo{journal}{Science (New York, N.Y.)}}
  \textbf{\bibinfo{volume}{275}}, \bibinfo{pages}{1767--1770}
  (\bibinfo{year}{1997}).
\newblock \urlprefix\url{http://www.ncbi.nlm.nih.gov/pubmed/9065395}.

\bibitem{CRC2017}
\bibinfo{author}{Choi, D.-J.} \emph{et~al.}
\newblock \bibinfo{title}{Mapping the orbital structure of impurity bound
  states in a superconductor}.
\newblock \emph{\bibinfo{journal}{Nature Communications}}
  \textbf{\bibinfo{volume}{8}}, \bibinfo{pages}{15175} (\bibinfo{year}{2017}).
\newblock \urlprefix\url{https://doi.org/10.1038/ncomms15175}.

\bibitem{PhysRevB.67.020502}
\bibinfo{author}{Morr, D.~K.} \& \bibinfo{author}{Stavropoulos, N.~A.}
\newblock \bibinfo{title}{Quantum interference between impurities: Creating
  novel many-body states in s-wave superconductors}.
\newblock \emph{\bibinfo{journal}{Phys. Rev. B}} \textbf{\bibinfo{volume}{67}},
  \bibinfo{pages}{020502} (\bibinfo{year}{2003}).
\newblock \urlprefix\url{https://link.aps.org/doi/10.1103/PhysRevB.67.020502}.

\bibitem{PhysRevB.73.224511}
\bibinfo{author}{Morr, D.~K.} \& \bibinfo{author}{Yoon, J.}
\newblock \bibinfo{title}{Impurities, quantum interference, and quantum phase
  transitions in $s$-wave superconductors}.
\newblock \emph{\bibinfo{journal}{Phys. Rev. B}} \textbf{\bibinfo{volume}{73}},
  \bibinfo{pages}{224511} (\bibinfo{year}{2006}).
\newblock \urlprefix\url{https://link.aps.org/doi/10.1103/PhysRevB.73.224511}.

\bibitem{doi:10.1126/science.1202204}
\bibinfo{author}{Franke, K.~J.}, \bibinfo{author}{Schulze, G.} \&
  \bibinfo{author}{Pascual, J.~I.}
\newblock \bibinfo{title}{Competition of superconducting phenomena and kondo
  screening at the nanoscale}.
\newblock \emph{\bibinfo{journal}{Science}} \textbf{\bibinfo{volume}{332}},
  \bibinfo{pages}{940--944} (\bibinfo{year}{2011}).
\newblock
  \urlprefix\url{https://www.science.org/doi/abs/10.1126/science.1202204}.
\newblock \eprint{https://www.science.org/doi/pdf/10.1126/science.1202204}.

\bibitem{PhysRevB.87.094518}
\bibinfo{author}{Stanescu, T.~D.}, \bibinfo{author}{Lutchyn, R.~M.} \&
  \bibinfo{author}{Das~Sarma, S.}
\newblock \bibinfo{title}{Dimensional crossover in spin-orbit-coupled
  semiconductor nanowires with induced superconducting pairing}.
\newblock \emph{\bibinfo{journal}{Phys. Rev. B}} \textbf{\bibinfo{volume}{87}},
  \bibinfo{pages}{094518} (\bibinfo{year}{2013}).
\newblock \urlprefix\url{https://link.aps.org/doi/10.1103/PhysRevB.87.094518}.

\bibitem{PhysRevB.102.174502}
\bibinfo{author}{Odobesko, A.} \emph{et~al.}
\newblock \bibinfo{title}{Anisotropic vortices on superconducting nb(110)}.
\newblock \emph{\bibinfo{journal}{Phys. Rev. B}}
  \textbf{\bibinfo{volume}{102}}, \bibinfo{pages}{174502}
  (\bibinfo{year}{2020}).
\newblock \urlprefix\url{https://link.aps.org/doi/10.1103/PhysRevB.102.174502}.

\bibitem{brinker2021anomalous}
\bibinfo{author}{Brinker, S.}, \bibinfo{author}{Küster, F.},
  \bibinfo{author}{Parkin, S. S.~P.}, \bibinfo{author}{Sessi, P.} \&
  \bibinfo{author}{Lounis, S.}
\newblock \bibinfo{title}{Anomalous excitations of atomically crafted quantum
  magnets}  (\bibinfo{year}{2021}).
\newblock \eprint{2111.02203}.

\bibitem{liebhaber2021quantum}
\bibinfo{author}{Liebhaber, E.} \emph{et~al.}
\newblock \bibinfo{title}{Quantum spins and hybridization in
  artificially-constructed chains of magnetic adatoms on a superconductor}
  (\bibinfo{year}{2021}).
\newblock \eprint{2107.06361}.

\bibitem{Odobesko.2019}
\bibinfo{author}{Odobesko, A.~B.} \emph{et~al.}
\newblock \bibinfo{title}{Preparation and electronic properties of clean
  superconducting nb(110) surfaces}.
\newblock \emph{\bibinfo{journal}{Physical Review B}}
  \textbf{\bibinfo{volume}{99}} (\bibinfo{year}{2019}).

\bibitem{Kuester2021}
\bibinfo{author}{K{\"u}ster, F.} \emph{et~al.}
\newblock \bibinfo{title}{Correlating {{Josephson}} supercurrents and {{Shiba}}
  states in quantum spins unconventionally coupled to superconductors}.
\newblock \emph{\bibinfo{journal}{Nature Communications}}
  \textbf{\bibinfo{volume}{12}}, \bibinfo{pages}{1108} (\bibinfo{year}{2021}).

\bibitem{Papanikolaou:2002}
\bibinfo{author}{Papanikolaou, N.}, \bibinfo{author}{Zeller, R.} \&
  \bibinfo{author}{Dederichs, P.~H.}
\newblock \bibinfo{title}{{Conceptual improvements of the KKR method}}.
\newblock \emph{\bibinfo{journal}{Journal of Physics: Condensed Matter}}
  \textbf{\bibinfo{volume}{14}}, \bibinfo{pages}{2799} (\bibinfo{year}{2002}).

\bibitem{Bauer:2014}
\bibinfo{author}{Bauer, D. S.~G.}
\newblock \bibinfo{title}{{Development of a relativistic full-potential
  first-principles multiple scattering {G}reen function method applied to
  complex magnetic textures of nanostructures at surfaces}}.
\newblock \emph{\bibinfo{journal}{Forschungszentrum J\"ulich}}
  (\bibinfo{year}{2014}).

\bibitem{Vosko:1980}
\bibinfo{author}{Vosko, S.~H.}, \bibinfo{author}{Wilk, L.} \&
  \bibinfo{author}{Nusair, M.}
\newblock \bibinfo{title}{{Accurate spin-dependent electron liquid correlation
  energies for local spin density calculations: a critical analysis}}.
\newblock \emph{\bibinfo{journal}{Canadian Journal of Physics}}
  \textbf{\bibinfo{volume}{58}}, \bibinfo{pages}{1200--1211}
  (\bibinfo{year}{1980}).

\bibitem{Liechtenstein1987}
\bibinfo{author}{Liechtenstein, A.~I.}, \bibinfo{author}{Katsnelson, M.~I.},
  \bibinfo{author}{Antropov, V.~P.} \& \bibinfo{author}{Gubanov, V.~A.}
\newblock \bibinfo{title}{Local spin density functional approach to the theory
  of exchange interactions in ferromagnetic metals and alloys}.
\newblock \emph{\bibinfo{journal}{Journal of Magnetism and Magnetic Materials}}
  \textbf{\bibinfo{volume}{67}}, \bibinfo{pages}{65--74}
  (\bibinfo{year}{1987}).

\bibitem{Ebert2009}
\bibinfo{author}{Ebert, H.} \& \bibinfo{author}{Mankovsky, S.}
\newblock \bibinfo{title}{Anisotropic exchange coupling in diluted magnetic
  semiconductors: {{Ab}} initio spin-density functional theory}.
\newblock \emph{\bibinfo{journal}{Physical Review B}}
  \textbf{\bibinfo{volume}{79}}, \bibinfo{pages}{045209}
  (\bibinfo{year}{2009}).

\bibitem{Brinker2019}
\bibinfo{author}{Brinker, S.}, \bibinfo{author}{Dias, M. d.~S.} \&
  \bibinfo{author}{Lounis, S.}
\newblock \bibinfo{title}{The chiral biquadratic pair interaction}.
\newblock \emph{\bibinfo{journal}{New Journal of Physics}}
  \textbf{\bibinfo{volume}{21}}, \bibinfo{pages}{083015}
  (\bibinfo{year}{2019}).

\end{thebibliography}



\section*{Supplementary material}

# Supplementary Information
# Non-Majorana modes in diluted spin chains proximitized to a superconductor

Küster *et al.*

# Supplementary Note 1: Metallic vs superconducting STM tip

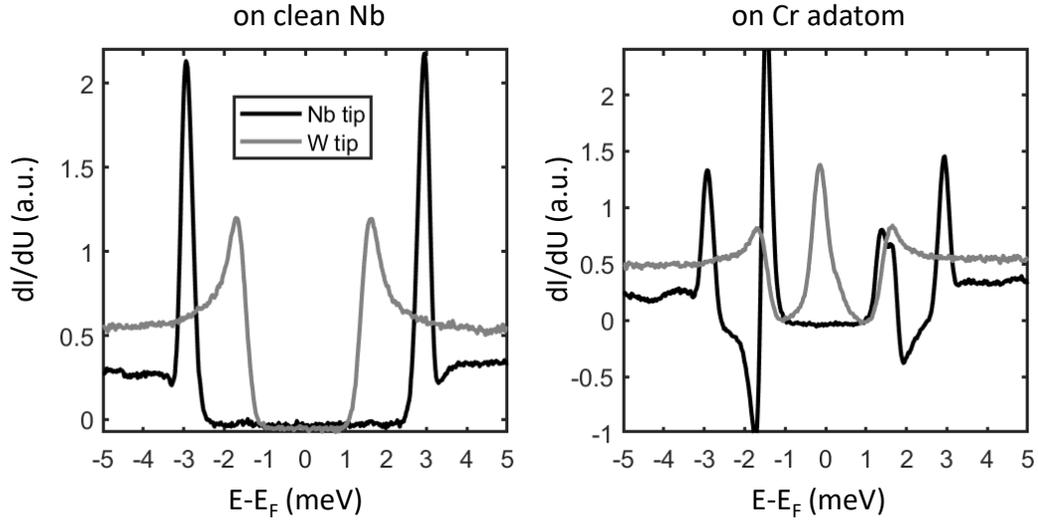

**Figure S1:** STS data taken with a normal metallic STM tip (grey) and a superconducting tip due to a Nb cluster at the apex obtained by tip indentations into the substrate. The left panel shows data on clean Nb. It shows that the gap size doubles when the tip is superconducting because of the convolution of tip and sample density of states, significantly enhancing the energy resolution. In the right panel depicting STS on a Cr adatom that features YSR resonance states inside the superconducting energy gap, the advantage of the enhanced energy resolution becomes clear since it enables a much more detailed analysis of the sharp in-gap peaks.

# Supplementary Note 2: Cr adatom - Orbital character of YSR states

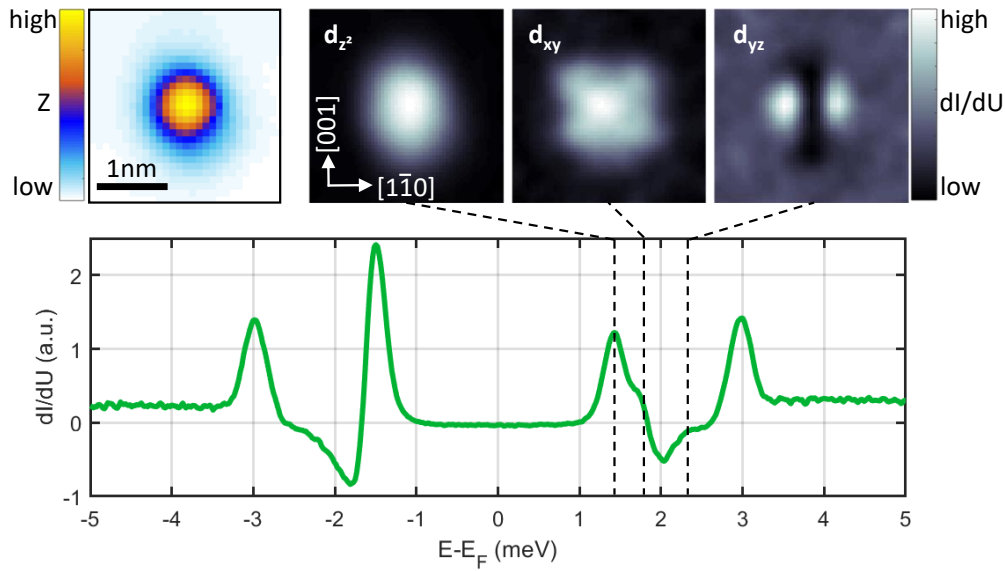

**Figure S2:** The figure shows spatial imaging of YSR resonances corresponding to different atomic orbitals. The topography data in the left panel has the same scale like the constant energy $dI/dU$ maps to the right, taken at energies 1.50 meV, 1.77 meV and 2.28 meV, from left to the right, indicated with dashed lines in the spectroscopy curve below taken on a single Cr adatom. Using STS, we observe the peak that corresponds to the Cr $d_{z^2}$ orbital as the most prominent one due to its longer extension into vacuum compared to the others.

# Supplementary Note 3: Longer chain

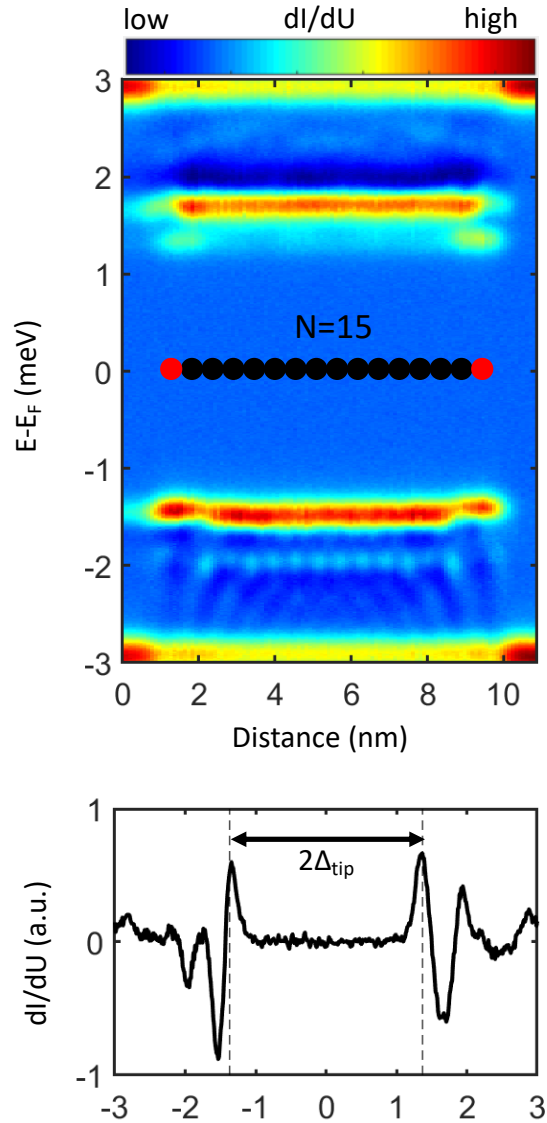

**Figure S3:** Spectroscopic mapping of the zero-energy mode appearing at boundary of a 15 adatom chain assembled along the $[1\bar{1}3]$ direction. The measurements have been acquired with a different microtip compared to the chain presented in the main text. The top panel report a full spectroscopy line across the chain. The middle panel highlights the rather symmetric electron- and hole- components of the zero-bias mode. To account for the asymmetric background at positive and negative energy (black lines in the top panel), the spectrum at the end of the chain has been normalized by subtracting the spectrum averaged over the bulk of the chain.

# Supplementary Note 4: Surface preparation

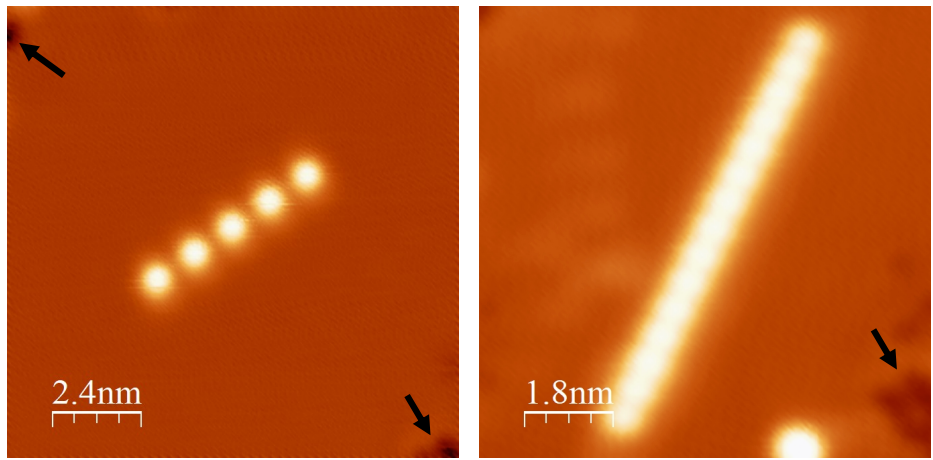

**Figure S4:** Topographic images showing the areas around Cr chains. The first image correspond to the area where all chains presented in Fig. 2 in the main text have been assembled and measured. The second image correspond to the longest chain along the [1$\bar{1}$3] direction. Because of the long range effect between Cr adatoms and the external environment, atomic manipulation techniques have been used to clean up the area before performing the experiments. The arrow indicate residual defects, all located some away from the chains, where their influence becomes negligible.

# Supplementary Note 5: Spectroscopic data for short chains.

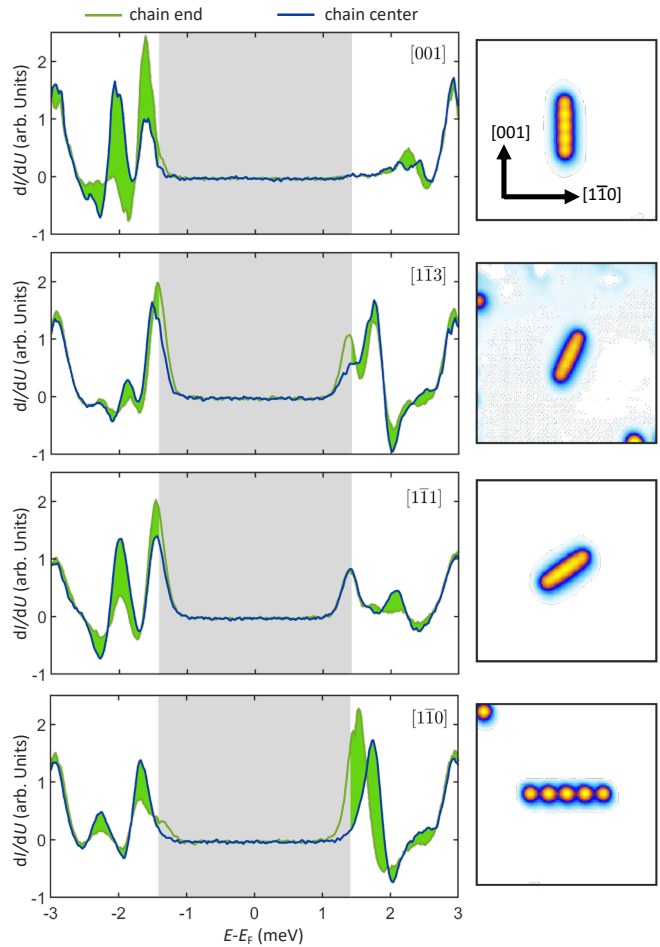

**Figure S5: Scanning tunneling spectroscopy data for short chains** Spectroscopic measurements acquired by positioning the tip at the center (blue line) and at the end (green line) for five adatoms chains built along distinct crystallographic directions. The gray area identifies the tip superconducting gap $\pm\Delta_{\text{tip}}$.

6

# Supplementary Note 6: Spectroscopic data for long chains.

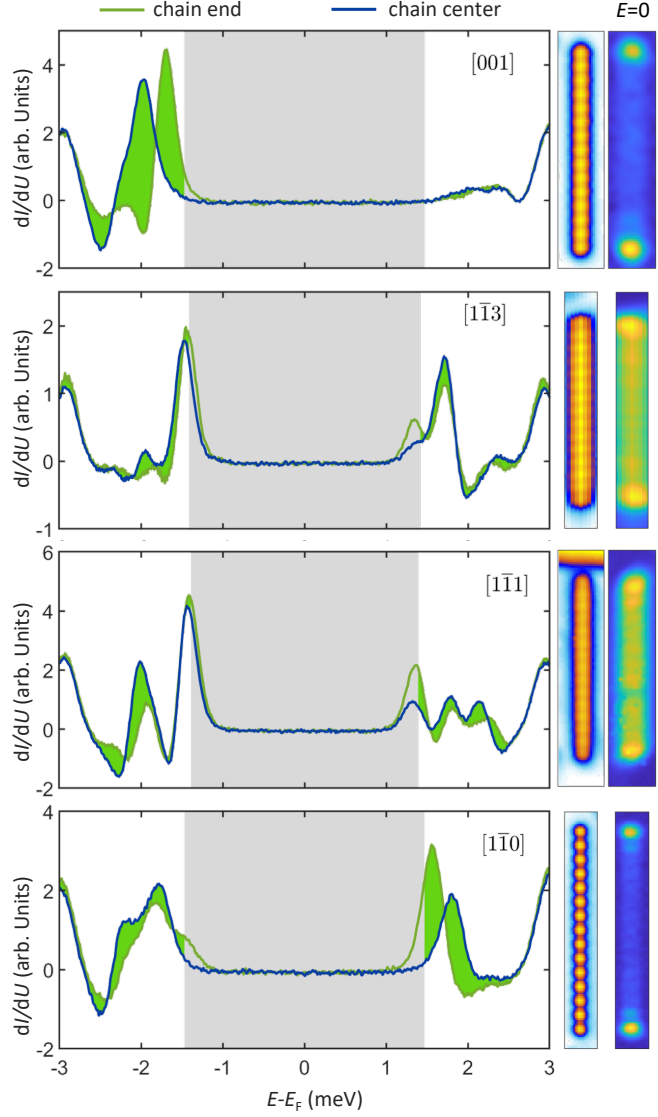

**Figure S6: Scanning tunneling spectroscopy data for long chains** Spectroscopic measurements acquired by positioning the tip at the center (blue line) and at the end of the chain (green line) for 15 adatoms chains built along distinct crystallographic directions. As for shorter chains (see Fig 2 in the main text) the maps acquired at zero energy highlight a significant spectral weight accumulation at the end of the chains. The gray area identifies the tip superconducting gap $\pm\Delta_{\text{tip}}$. The zero energy maps have been obtained by considering the tip superconducting gap, i.e. sum of the intensities at $\pm\Delta_{\text{tip}}$

7

# Supplementary Note 7: Boundary modes in the metallic regime.

Although being in the diluted regime, where abrupt structural changes at the edge are negligible, the existence of boundary modes localized at zero energy is confirmed by studying the same chains discussed in Fig. 2 once driving the system into the normal metallic regime, where Cr adatoms show a clear zero bias spectroscopic signature [1]. These results are illustrated in Figure S7. Its inspection reveals, also in the metallic regime, the existence of boundary modes which make the end of the chain distinct with respect to the center. This is evident by comparing the zero bias anomaly acquired by positioning the tip on each one of the adatom (see Fig.S7 d-f) and further highlighted in panels where spectra acquired in the center of the chains are directly overlapped to the spectra acquired on the last adatom (see Fig.S7g-i). The localization of this boundary modes takes place on the same length scale as in the superconducting regime, i.e. the single atom level.

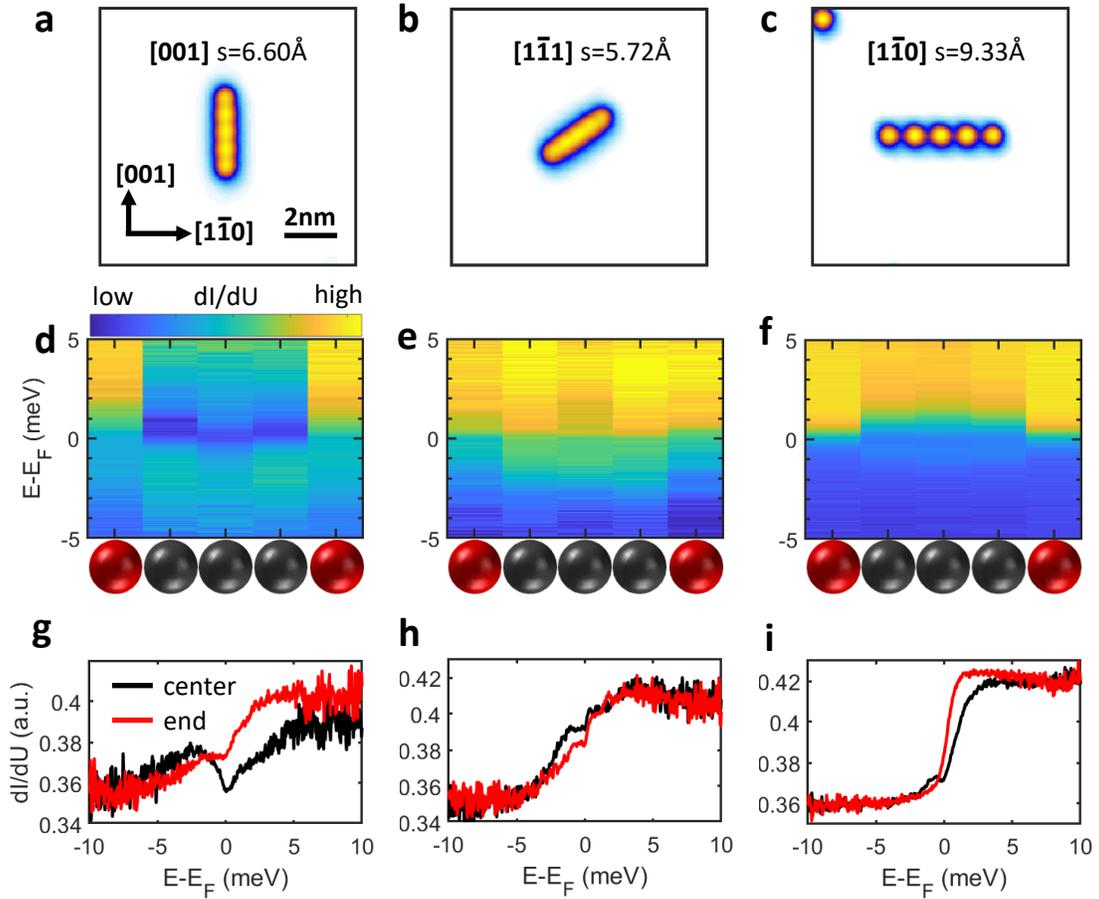

Figure S7: **Distinct boundary behaviour in the metallic regime.** **a-c** Topographic images five adatoms chains oriented along distinct crystallographic directions. **d-f** Spectroscopic mapping of the chain in the metallic regime, achieved by the application of a magnetic field of 1T perpendicular to the sample surface. A significant change in the zero bias anomaly is evident at the chain end, independently from their different crystallographic direction. These differences are highlighted in panels **g-i**.

# Supplementary Note 8: Minimal tight-binding model

## Supplementary Note 8.1: Tight-binding model

Here, we present the tight-binding model used in Sec. **Origin of zero energy boundary modes** of the main manuscript: the impurity chain, embedded in a two-dimensional superconductor, is described by the Hamiltonian

$$H_0 = \sum_{n,m} \left[ \sum_{\nu,\nu'} c^\dagger_{n,m,\nu} \left\{ (4t-\mu)\delta_{\nu,\nu'} + J_{n,m}(\sigma_z)_{\nu,\nu'} \right\} c_{n,m,\nu'} - \left\{ \sum_{\nu,\nu'} (t c^\dagger_{n,m,\nu} c_{n+1,m,\nu'} + t c^\dagger_{n,m,\nu} c_{n,m+1,\nu'}) \delta_{\nu,\nu'} \right. \right.$$
$$\left. \left. + \Delta c^\dagger_{n,m,\downarrow} c^\dagger_{n,m,\uparrow} + \text{H.c.} \right\} \right], \tag{S.1}$$

where $t$ and $\mu$ denote the matrix hopping element and the chemical potential, respectively. Furthermore, $\sigma_z$ is the Pauli z matrix acting in spin space and $\delta_{\nu,\nu'}$ is the Kronecker delta. The operator $c^\dagger_{n,m,\nu}$ ($c_{n,m,\nu}$) creates (annihilates) an electron with spin $\nu$ at the site $(n,m)$, where n and m denote the $x$ and $y$ position of the site in the two-dimensional lattice, respectively. The total number of lattice sites is given by $N_x \times N_y$. In addition, $J_{n,m}$ denotes the exchange coupling between the magnetic moments of the impurities and the spin of the itinerant electrons at the site $(n,m)$; and $\Delta$ is the superconducting gap. We set the lattice spacing $a$ to one. The system described by the Hamiltonian $H_0$ cannot enter a topological phase, except when Rashba spin-orbit interaction (SOI) is also present in the system. Such a Rashba SOI term is modelled by

$$H_R = \alpha \sum_{n,m} \left[ \left( c^\dagger_{n,m,\uparrow} c_{n+1,m,\downarrow} - c^\dagger_{n,m,\downarrow} c_{n+1,m,\uparrow} \right) - i \left( c^\dagger_{n,m,\uparrow} c_{n,m+1,\downarrow} + c^\dagger_{n,m,\downarrow} c_{n,m+1,\uparrow} \right) + \text{H.c.} \right], \tag{S.2}$$

where $\alpha$ denotes the Rashba SOI strength. We apply open boundary conditions and diagonalize the matrix $H = H_0 + H_R$, this yields the eigenvalues $E^l$ and the associated eigenvectors $\Psi^l$ consisting of the well known coherence factors $u^l_{n,m,\nu}$ and $v^l_{n,m,\nu}$. Furthermore, we define the local density of states (LDOS) at the site $(n,m)$ and at energy $\omega$ as

$$\rho(\omega, n, m) = -\frac{1}{\pi} \text{Im} \left[ \sum_{l,\nu} \left( \frac{|u^l_{n,m,\nu}|^2}{\omega - E_l + i\epsilon} + \frac{|v^l_{n,m,\nu}|^2}{\omega + E_l + i\epsilon} \right) \right], \tag{S.3}$$

where the sum runs over the positive eigenvalues, and the parameter $\epsilon$ accounts for a broadening of the energy levels. We ignore states $\Psi^l$ with energies $E^l > \Delta$ for the calculation of $\rho$, since these states are energetically well separated from the lowest sub-gap states and we are mainly interested in the LDOS close to zero-energy.

## Supplementary Note 8.2: Topological regime

In Sec. **Origin of zero energy boundary modes** of the main manuscript we focus on the trivial regime by neglecting SOI. Here, in the supplementary note we take SOI into account; the model can therefore enter the topological phase and could, in principle, host Majorana bound states (MBSs) under certain conditions. Fig. **S8**a shows the energies of a two-dimensional superconductor with a few adatoms, ordered as nearest neighbours, as a function of the $s-d$ exchange coupling strength $J$. The gap closing and reopening, which is predicted to accompany the topological phase transition, is only weakly pronounced due to the limited small number of adatoms. Furthermore, the restricted size of the system does not admit stable zero-energy states due to a strong overlap and hybridization of the MBSs. The results are slightly different in case of indirectly coupled adatoms, see Fig. **S8**b. The system exhibits a stable zero-energy state for some range of the exchange coupling. Moreover, the zero-energy state disappears at the exchange coupling $J = J_1$, where

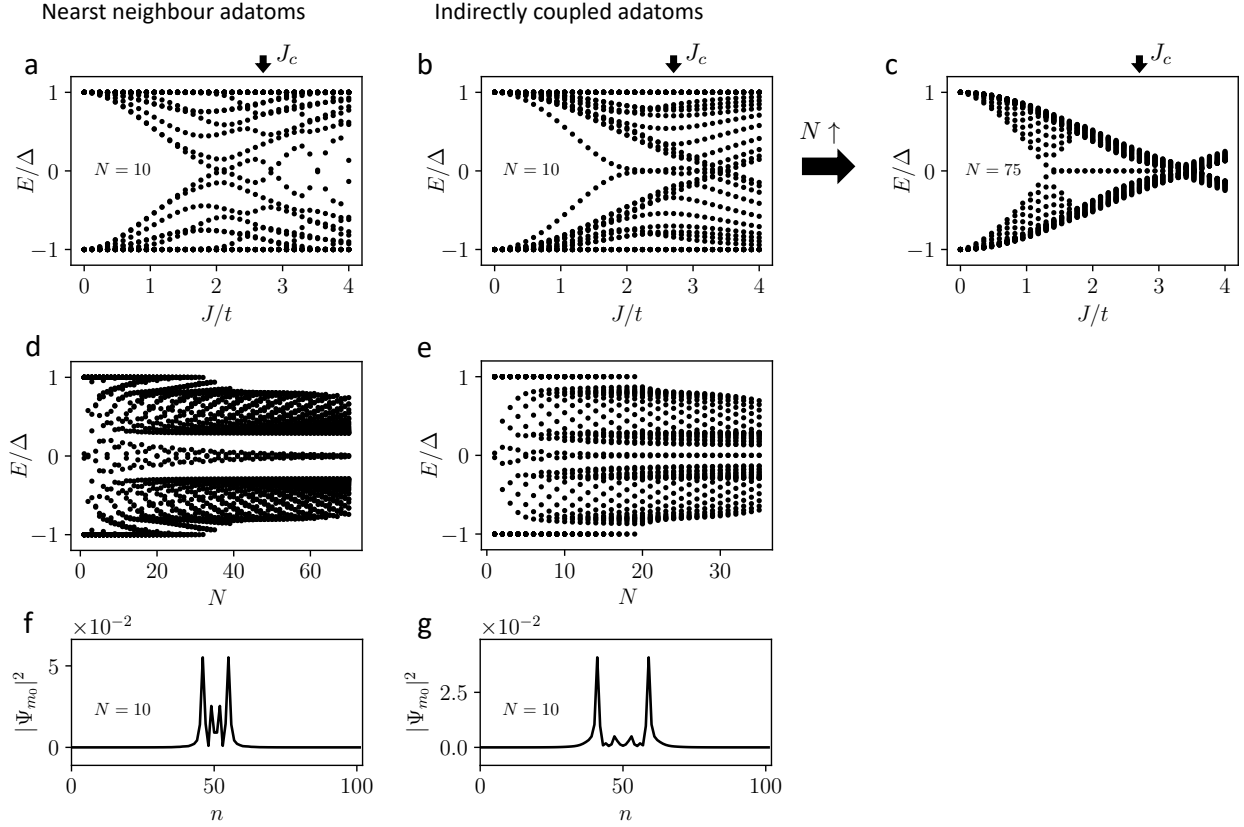

**Supplementary Figure S8: Energies and probability densities of systems including SOI.** (**a, b, c**) Energies $E$ as a function of the $s-d$ exchange coupling $J$. The black arrows above the panels indicate the critical exchange coupling $J_c$ for which a single bound state assumes zero energy. (**d, e**) Lowest energies as a function of the number of impurities $N$ at $J = J_c$. (**f, g**) Probability densities of the lowest state at $J = J_c$ for a system with $N = 10$ impurities. The parameters used in the plots are: $t = 1$, $\mu = 1$, $\alpha = 0.5$, $\Delta = 0.2$ and $J_c \approx 2.71$. Furthermore, we choose the lattice size $N_x \times N_y = 102 \times 60$, except in panel **c** in which we set $N_x \times N_y = 202 \times 50$. The chain is deposited at the $y$ coordinate $m_0 = 31$, except in panel **c**, where we set $m_0 = 26$. The impurities are separated by $L = a$ ($L = 2a$) in the panels **a**, **d**, and **f** (panels **b**, **c**, **e**, and **g**)

the gap closes and reopens. This closing and reopening, however, does not indicate a phase transition from trivial to topological. A second bulk gap closing and reopening, associated with this phase transition, appears only in much longer chains approximately at the exchange coupling for which the zero-energy pinning of the energetically lowest state starts, see Fig. S8c. In addition, in long chains it turns out that the gap closing and reopening at $J_1$ initializes a topological phase transition from topological to trivial rather than from trivial to topological phase.

In order to compare the numerical model with the experiment, in particular, with the data shown in Fig. **1**b, we repeat the procedure described in the main manuscript: First, we consider a single magnetic impurity and calculate the energy as a function of the exchange coupling. The energy of the sub-gap state bound to the impurity is zero for $J = J_c$, indicated by the black arrow in Figs. S8a-c. Finally, we fix the exchange coupling to $J = J_c$ and calculate the energies as a function of the number of adatoms $N$, see Fig. S8d. The energies oscillate as a function of $N$ and the amplitude of these oscillations decays for values of $N$, which are much larger than the number of adatoms in the experiment, and a MBS stabilizes (one at

each end of the chain). Moreover, the MBS is energetically separated from the bulk states of the chain by a topological minigap $\Delta_T$. This minigap sets, among other parameters, the spatial localization of the MBS. A small minigap leads to spatially extended MBSs from each chain end which overlap and hybridize and form a non-zero energy fermionic bound state. Consequently, the formation of true MBSs in a system with a small minigap requires long chains. The experimental data, see Figs. **1**b and **1**d, reveal only a minor energetic separation between bulk and the edge states. It is therefore unlikely that the combination of a small minigap and a small number of adatoms support MBSs. Furthermore, the bulk modes of the chain are restricted in the experiment to a narrow energy interval, in theory, however, these states can assume energies $E_{Bulk}$ in the range $\Delta_T \leq E_{Bulk} < \Delta$, unless the Shiba bound states from the different adatoms are very weakly coupled. However, in the limit of almost uncoupled sub-gap states, the parameter space for the topological phase shrinks almost to zero. Finally, the oscillations of the lowest energy as a function of $N$ have not been observed in the experiment. Consequently, a topological origin of the zero-energy edge modes is unlikely. The experimental platform itself, however, represents an excellent starting point for the study of MBSs. Increasing the number of magnetic impurities might lead to further signatures, beside the zero-energy peak and its spectral weight on the ends of the chain, indicating the presence of MBSs. In addition, we numerically study the probability density of the lowest state in a short chain. The lowest state which transforms with growing chain length into MBSs has already higher weights on the ends of the chain, see Fig. **S8**f. For the sake of completeness, we also present the case of a chain with indirectly coupled adatoms, see Fig. **S8**e. The topological minigap is in this case much smaller, which agrees better with the energetic separation of end and bulk modes observed in the experiment. The energetic restriction to a narrow energy interval, however, still does not match with the numerical calculation. The probability density of a chain with ten impurities reveals again strong peaks at the ends of the chain, since the chain is effectively longer than the one build of nearest neighbour adatoms, see Fig. **S8**g. Finally, we conclude that only sufficiently long chains support the formation of topologically protected MBSs. This means, in particular, that the chain length needs to be much larger than the localization length of MBSs [2, 3].

## Supplementary Note 9:  Magnetic ground-state properties

To describe the magnetic structure of the Cr chains we use a generalized Heisenberg model,

$$H = \sum_i \mathbf{e}_i \mathcal{K}_i \mathbf{e}_i + \frac{1}{2} \sum_{ij} J_{ij} \mathbf{e}_i \cdot \mathbf{e}_j + \frac{1}{2} \sum_{ij} \mathbf{D}_{ij} \cdot (\mathbf{e}_i \times \mathbf{e}_j) + \frac{1}{2} \sum_{ij} \mathbf{e}_i J_{ij}^{\text{aniso}} \mathbf{e}_j \quad . \tag{S.4}$$

where $i$ and $j$ label the spins with a magnetic moment $\mathbf{m}_i = m_i \mathbf{e}_i$, $\mathcal{K}_i$ is the magnetic on-site anisotropy, $J_{ij}$ is the istropic exchange interaction, $\mathbf{D}_{ij}$ is the Dzyaloshinksii-Moriya interaction, and $J_{ij}^{\text{aniso}}$ is the symmetric anisotropic exchange. We assume that the on-site anisotropy is similar to that of the isolated Cr adatom,

$$\mathcal{K} = \begin{pmatrix} -0.02 & 0 & 0 \\ 0 & -0.17 & 0 \\ 0 & 0 & 0.19 \end{pmatrix} \text{meV} \quad , \tag{S.5}$$

which was obtained from the method of constraining fields [4]. This assumption was tested by using the [001] chain as an example, for which we do not find any significant deviations from the isolated adatom values. The magnetic exchange interactions $J_{ij}$ and $\mathbf{D}_{ij}$ of nearest neighbors and next-nearest neighbors are shown in Supplementary Figure **S9**a and b, respectively. As discussed in the main manuscript there are no significant edge effects observable, except for a small increase of the ferromagnetic coupling in case of the [1$\bar{1}$1] chain.

The magnetic structure is obtained by minimizing the Heisenberg model, which results in the angles shown in Figure 3 of the main manuscript and Supplementary Table **S1**. Due to the weakness of the DMI there is no significant non-collinearity present in the chains.

|        | [001] $\vartheta$ | [001] $\varphi$ | [1$\bar{1}$3] $\vartheta$ | [1$\bar{1}$3] $\varphi$ | [1$\bar{1}$1] $\vartheta$ | [1$\bar{1}$1] $\varphi$ | [1$\bar{1}$0] $\vartheta$ | [1$\bar{1}$0] $\varphi$ |
|--------|-------|-------|-------|-------|-------|-------|-------|-------|
| Atom 1 | 89.1° | 90.0° | 91.8° | 260.5° | 95.0° | 268.8° | 90.0° | 270.0° |
| Atom 2 | 90.5° | 270.0° | 89.9° | 260.2° | 92.2° | 269.4° | 90.0° | 90.0° |
| Atom 3 | 90.0° | 90.0° | 90.0° | 260.2° | 90.0° | 269.4° | 90.0° | 270.0° |
| Atom 4 | 89.6° | 270.0° | 90.1° | 260.2° | 87.8° | 269.4° | 90.0° | 90.0° |
| Atom 5 | 90.9° | 90.0° | 88.2° | 260.5° | 85.0° | 268.8° | 90.0° | 270.0° |

**Supplementary Table S1:** Magnetic ground states obtained from minimizing a Heisenberg model, eq. (S.4), with parameters shown in Supplementary Figure **S9** and eq. (S.5).

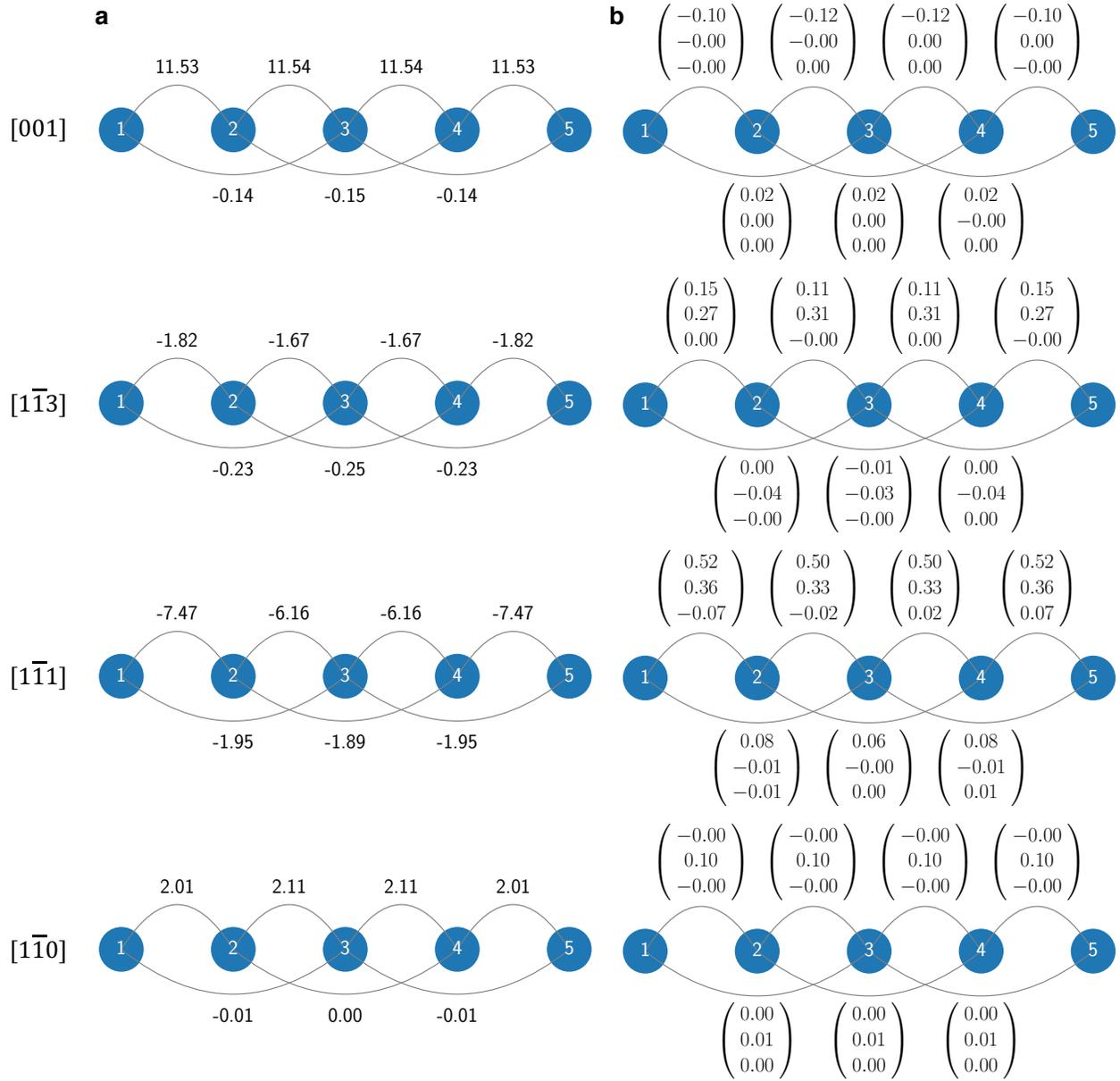

**Supplementary Figure S9:** Magnetic exchange interactions of the investigated 5-atomic chains along the [001], [1$\bar{1}$3], [1$\bar{1}$1], and [1$\bar{1}$0] directions. Shown are the nearest and next-nearest neighboring isotropic exchange interactions $J$ (**a**) and Dzyaloshinskii-Moriya interactions **D** (**b**). The magnetic ground states obtained from these parameters are shown in Supplementary Table **S1**. The coordinate frame of the DMI vector is defined by the [1$\bar{1}$0], [001], and [110] directions.

13

# Supplementary Note 10: Multi-orbital tight-binding model and topological invariant

In the following, we use a tight-binding model with first-principles input to describe properties of the magnetic chains in the superconducting regime. We used the procedure in several works and refer the reader also to References [5, 6, 1], for additional details. Using the inverse of the Green function of the different chains obtained from first principles, we construct the hybridization function of the chain complexes

$$\mathcal{H} = E - G^{-1}(E) \quad , \tag{S.6}$$

which contains various information such as the hybridization strength with the surface, the crystal field splitting, the strength of spin-orbit coupling, and the strength of the direct hopping between the chain atoms. The hamiltonian can be written as an effective tight-binding model, where the on-site part is given by (omitting the atom index $i$ on the parameters)

$$\mathcal{H}_i = \sum_{mm'} \sum_{ss'} \big( E_d\,\delta_{mm'}\delta_{ss'} + U\,\mathbf{e}\cdot\boldsymbol{\sigma}_{ss'}\delta_{mm'} + \lambda\,\mathbf{L}_{mm'}\cdot\boldsymbol{\sigma}_{ss'} + \Delta^{(\mathrm{re})}_{mm'}\delta_{ss'}$$
$$+ \mathrm{i}\Gamma\,\delta_{mm'}\delta_{ss'} + \mathrm{i}\Delta^{(\mathrm{im})}_{mm'}\delta_{ss'}\big)c^\dagger_{ims}c_{im's'} \quad , \tag{S.7}$$

and the inter-atomic hopping part is given by

$$\mathcal{H}_{ij} = \sum_{mm'} \sum_{s} t_{mm'} c^\dagger_{ims} c_{jm's} \quad . \tag{S.8}$$

$E_d$ is the average energy of the $d$-orbitals with respect to the Fermi energy, $2U$ represents the exchange splitting of the magnetic moment pointing along $\mathbf{e}$, $\boldsymbol{\sigma} = (\sigma_x, \sigma_y, \sigma_z)$ is the vector of Pauli matrices, $\lambda$ is the strength of the local spin-orbit coupling (-varies between 7 to 10 meV across the chains and directions), $\mathbf{L}$ is the local orbital angular momentum operator, $\Delta^{(\mathrm{re})}$ is an orbital dependent energy shift corresponding to the crystal field splitting, $\Gamma$ and $\Delta^{(\mathrm{im})}$ are non-hermitian contributions that result from the hybridization with the substrate, and $t_{mm'}$ is the orbital-dependent hopping between atoms $i$ and $j$. The crystal field splitting $\Delta^{(\mathrm{re})}$ of an isolated Cr adatom is shown in Supplementary Table **S2**, the nearest neighbor hoppings of all considered chains are shown in Supplementary Table **S3**, and more long-ranged hoppings in the [001] chain are shown in Supplementary Table **S4**.

We model the impact of the proximity-induced superconductivity on the magnetic chains by a $s$-wave pairing using the following form of the Bogouliubov-de-Gennes Hamiltonian,

$$\mathcal{H}^{\mathrm{BdG}} = \begin{pmatrix} \mathcal{H}^{\mathrm{chain}} & \Delta \mathbf{1}_{mm'} \otimes \mathrm{i}\sigma^y \\ -\Delta \mathbf{1}_{mm'} \otimes \mathrm{i}\sigma^y & -(\mathcal{H}^{\mathrm{chain}})^{\mathrm{T}} \end{pmatrix} \quad , \tag{S.9}$$

with $\mathcal{H}^{\mathrm{chain}} = \sum_i \mathcal{H}_i + \sum_{ij} \mathcal{H}_{ij}$.

From eqs. (S.7)-(S.9), we can directly define the Hamiltonian of an infinite one-dimensional chain in $k$-space using the Fourier transform,

$$\mathcal{H}^{\mathrm{BdG}}_{\mu\nu}(k) = \sum_j \exp(\mathrm{i}k(a_{ij} + R_{\mu\nu}))\mathcal{H}^{\mathrm{BdG}}_{ij,\mu\nu} \quad , \tag{S.10}$$

where the index $i$ was left in for the sake of clarity and has no relevance for the infinite chain due to the translational invariance, $\mu$ and $\nu$ label the sites within a unit cell containing $N_{\mathrm{uc}}$ atoms, and $a_{ij}$ and $R_{\mu\nu}$ are the basis vectors. To simplify the later discussion we also assume that the lattice constant can be set to 1 resulting in $a_{ij} = (j-i)N_{\mathrm{uc}}$, $R_{\mu\nu} = \mu - \nu$ and $k \in [-\pi/N_{\mathrm{uc}}, \pi/N_{\mathrm{uc}}]$.

## Supplementary Note 10.1: YSR states of the isolated Cr adatom

In this section, we revisit the case of a single Cr adatom, which was extensively discussed in the Supplementary Notes of Ref. [5], to which we also refer for more details. Using the effective Hamiltonian, eq. (S.7),

gives direct access to the scattering phase shifts of each orbital $\delta_m^\pm$ and the related energies of the YSR states $\epsilon_m$,

$$\frac{\epsilon}{\Delta} = \pm \cos\left(\delta_m^+ - \delta_m^-\right) \quad , \tag{S.11}$$

where $\Delta$ is the superconducting gap. In the case of the Cr adatom the YSR energies can be cast into the form

$$\frac{\epsilon_m}{\Delta} \approx \frac{1 - \alpha_m^2}{1 + \alpha_m^2} \quad \text{with} \quad \alpha_m = \frac{\Gamma_m U}{(E_m + U)(E_m - U)} \quad . \tag{S.12}$$

| $\Delta_{mm'}^{(\text{re})}$ | $xy$ | $yz$ | $z^2$ | $xz$ | $x^2 - y^2$ |
|---|---|---|---|---|---|
| $xy$ | -0.07 | 0.00 | 0.00 | 0.00 | 0.00 |
| $yz$ | 0.00 | -0.08 | 0.00 | 0.00 | 0.00 |
| $z^2$ | 0.00 | 0.00 | 0.39 | 0.00 | -0.20 |
| $xz$ | 0.00 | 0.00 | 0.00 | -0.04 | 0.00 |
| $x^2 - y^2$ | 0.00 | 0.00 | -0.20 | 0.00 | -0.19 |

**Supplementary Table S2:** Crystal field splitting of the isolated Cr adatom in unit of [eV].

| [001] | $xy$ | $yz$ | $z^2$ | $xz$ | $x^2 - y^2$ |
|---|---|---|---|---|---|
| $xy$ | -0.59 | 0.00 | 0.00 | 0.00 | 0.00 |
| $yz$ | 0.00 | 0.03 | 0.00 | 0.00 | 0.00 |
| $z^2$ | 0.00 | 0.00 | 0.10 | 0.00 | -0.01 |
| $xz$ | 0.00 | 0.00 | 0.00 | -0.02 | 0.00 |
| $x^2 - y^2$ | 0.00 | 0.00 | -0.01 | 0.00 | -0.16 |
| [$1\bar{1}3$] | $xy$ | $yz$ | $z^2$ | $xz$ | $x^2 - y^2$ |
| $xy$ | -0.10 | 0.00 | 0.17 | 0.00 | 0.09 |
| $yz$ | 0.00 | -0.13 | 0.00 | 0.08 | 0.00 |
| $z^2$ | 0.17 | 0.00 | 0.15 | 0.00 | 0.01 |
| $xz$ | 0.00 | 0.08 | 0.00 | -0.25 | 0.00 |
| $x^2 - y^2$ | 0.09 | 0.00 | 0.01 | 0.00 | 0.05 |
| [$1\bar{1}1$] | $xy$ | $yz$ | $z^2$ | $xz$ | $x^2 - y^2$ |
| $xy$ | 0.32 | -0.00 | 0.02 | 0.00 | -0.02 |
| $yz$ | 0.00 | 0.03 | 0.00 | -0.03 | 0.00 |
| $z^2$ | 0.02 | 0.00 | 0.08 | -0.01 | 0.10 |
| $xz$ | 0.00 | -0.03 | -0.01 | 0.14 | 0.00 |
| $x^2 - y^2$ | -0.02 | 0.00 | 0.10 | 0.00 | 0.07 |
| [$1\bar{1}0$] | $xy$ | $yz$ | $z^2$ | $xz$ | $x^2 - y^2$ |
| $xy$ | -0.25 | 0.00 | 0.00 | 0.0 | 0.00 |
| $yz$ | 0.00 | -0.03 | 0.00 | 0.0 | 0.00 |
| $z^2$ | 0.00 | 0.00 | 0.15 | 0.0 | -0.01 |
| $xz$ | 0.00 | 0.00 | 0.00 | 0.1 | 0.00 |
| $x^2 - y^2$ | 0.00 | 0.00 | -0.01 | 0.0 | -0.05 |

**Supplementary Table S3:** Hopping matrices for the chains along [001], [$1\bar{1}3$], [$1\bar{1}1$], and [$1\bar{1}0$] directions. Shown are the nearest-neighbor Cr-Cr hopping matrices in unit of [eV].

| $t_{13}$ | $xy$ | $yz$ | $z^2$ | $xz$ | $x^2-y^2$ |
|---|---|---|---|---|---|
| $xy$ | 0.22 | 0.00 | 0.00 | 0.00 | 0.00 |
| $yz$ | 0.00 | 0.03 | 0.00 | 0.00 | 0.00 |
| $z^2$ | 0.00 | 0.00 | 0.14 | -0.00 | 0.02 |
| $xz$ | 0.00 | 0.00 | -0.00 | 0.04 | 0.00 |
| $x^2-y^2$ | 0.00 | 0.00 | 0.02 | 0.00 | -0.03 |
| $t_{14}$ | $xy$ | $yz$ | $z^2$ | $xz$ | $x^2-y^2$ |
| $xy$ | -0.16 | 0.00 | 0.00 | 0.00 | 0.00 |
| $yz$ | 0.00 | 0.00 | 0.00 | 0.00 | 0.00 |
| $z^2$ | 0.00 | 0.00 | -0.12 | 0.00 | -0.02 |
| $xz$ | 0.00 | 0.00 | 0.00 | -0.04 | 0.00 |
| $x^2-y^2$ | 0.00 | 0.00 | -0.02 | 0.00 | -0.01 |
| $t_{15}$ | $xy$ | $yz$ | $z^2$ | $xz$ | $x^2-y^2$ |
| $xy$ | 0.15 | 0.00 | 0.00 | 0.00 | 0.00 |
| $yz$ | 0.00 | -0.04 | 0.00 | 0.00 | 0.00 |
| $z^2$ | 0.00 | 0.00 | 0.07 | 0.00 | 0.01 |
| $xz$ | 0.00 | 0.00 | 0.00 | 0.02 | 0.00 |
| $x^2-y^2$ | 0.00 | 0.00 | 0.01 | 0.00 | -0.01 |

**Supplementary Table S4:** Hopping matrices for the [001] chain and the second, the third and the fourth neighbor hopping in units of [eV].

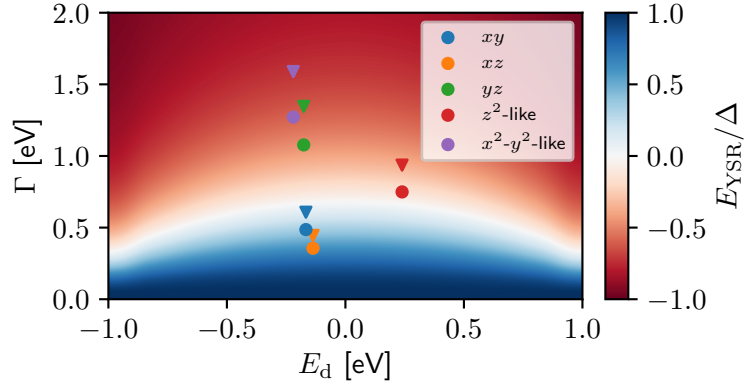

**Supplementary Figure S10:** Energy of the YSR states $E_{\text{YSR}}/\Delta$ of the isolated Cr adatom as function of the energy of the orbital $E_d$ and its hybridization strength $\Gamma$. Shown are the full parametrizations obtained from first principles (triangles), as well as the energies obtained from a reduced hybridization $\tilde{\Gamma} = 0.8\,\Gamma$ (circles).

The YSR energies obtained from this scheme using the first-principles parameters are shown as triangles in Supplementary Figure S10. To match the experimentally observed energy spectrum of the YSR states of Ref. [5], we renormalize the hybridization $\Gamma$ by a factor of 0.8, which is shown by the circles in Supplementary Figure S10. With this approach, we ensure that we obtain the $z^2$-like YSR state near zero energy, and we also obtain the correct order of the experimentally observed YSR states. The need for this renormalization can have several origins, e.g. an overestimation of the geometric relaxation of the adatom towards the surface could increase the hybridization or the general parameterization could be renormalized in the superconducting regime, which we do not address from first principles.

## Supplementary Note 10.2: Magnetism and topological invariant

For the classification of the topology of the used tight-binding model, special care has to be taken with respect to the non-hermicity. A detailed discussion of the topology in non-hermitian physics can be found, for example, in Ref. [7]. The particle-hole symmetry of the non-hermitian Hamiltonian defined in eq. (S.7) is reflected as

$$\mathcal{C}_-^{-1} H^\mathrm{T}(k) \mathcal{C}_- = -H(-k) \qquad \text{with} \qquad \mathcal{C}_- \mathcal{C}_-^* = +1 \quad . \tag{S.13}$$

$\mathcal{C}_-$ is given by $\tau_x = \begin{pmatrix} 0 & 1 \\ 1 & 0 \end{pmatrix}$ acting in the particle-hole space. For this case, a $\mathbb{Z}_2$ invariant $\nu \in \{0, 1\}$ can be defined by

$$(-1)^\nu = \frac{\mathrm{Pf}\left[H(\pi)\mathcal{C}_-\right]}{\mathrm{Pf}\left[H(0)\mathcal{C}_-\right]} \exp\left[-\frac{1}{2} \int_0^\pi \mathrm{d}k \log \det \left[H(k)\mathcal{C}_-\right]\right] \quad , \tag{S.14}$$

where Pf denotes the Pfaffian. This definition for the topological invariant is used in the phase diagrams shown in Figure 5b of the main manuscript.

# References

[1] Küster, F., Brinker, S., Lounis, S., Parkin, S. S. P. & Sessi, P. Long range and highly tunable coupling between local spins coupled to a superconducting condensate (2021). 2106.14932.

[2] Rainis, D., Trifunovic, L., Klinovaja, J. & Loss, D. Towards a realistic transport modeling in a superconducting nanowire with majorana fermions. *Phys. Rev. B* **87**, 024515 (2013). URL https://link.aps.org/doi/10.1103/PhysRevB.87.024515.

[3] Zyuzin, A. A., Rainis, D., Klinovaja, J. & Loss, D. Correlations between majorana fermions through a superconductor. *Phys. Rev. Lett.* **111**, 056802 (2013). URL https://link.aps.org/doi/10.1103/PhysRevLett.111.056802.

[4] Brinker, S., Dias, M. d. S. & Lounis, S. The chiral biquadratic pair interaction. *New Journal of Physics* **21**, 083015 (2019).

[5] Küster, F. *et al.* Correlating Josephson supercurrents and Shiba states in quantum spins unconventionally coupled to superconductors. *Nature Communications* **12**, 1108 (2021).

[6] Schneider, L. *et al.* Controlling in-gap end states by linking nonmagnetic atoms and artificially-constructed spin chains on superconductors. *Nature Communications* **11**, 4707 (2020). URL https://doi.org/10.1038/s41467-020-18540-3.

[7] Kawabata, K., Shiozaki, K., Ueda, M. & Sato, M. Symmetry and Topology in Non-Hermitian Physics. *Physical Review X* **9**, 041015 (2019).



\end{document}